\documentclass[11pt,xcolor=dvipsnames]{article}
\pdfoutput=1 
\usepackage[utf8x]{inputenc}
\usepackage[english]{babel}
\usepackage{graphicx}           
\usepackage{url}
\usepackage{eurosym}
\usepackage{makeidx}
\usepackage[plain]{algorithm}
\usepackage{algorithmic} 
\usepackage{color} 
\usepackage[margin=3.3cm]{geometry}
\usepackage{setspace}
\usepackage{datetime}
\usepackage{pifont}
\usepackage{rotating}
\usepackage{tikz}
\usepackage{natbib}
\usepackage{pstricks}
\usepackage{hyperref}
\usepackage{lscape}
\usepackage{subcaption}
\usepackage{amsmath, amsthm,amsfonts,amssymb,latexsym,stmaryrd}

\renewcommand{\tilde}{\widetilde}

\begin{document}
\title{Identifying conversion efficiency as a key mechanism underlying food webs adaptive  evolution: A step forward, or backward?}
\author{Coralie Fritsch$^{1}$ \and Sylvain Billiard$^{2}$ \and Nicolas Champagnat$^{1}$}

\footnotetext[1]{Universit\'e de Lorraine, CNRS, Inria, IECL, F-54000 Nancy, France}
\footnotetext[2]{Universit\'e de Lille, CNRS, UMR 8198 – Evo-Eco-Paleo, F-59000 Lille, France  \protect \\ 
				E-mail: {coralie.fritsch@inria.fr}, {sylvain.billiard@univ-lille.fr}, {nicolas.champagnat@inria.fr}.
				 }
\maketitle

\begin{abstract}
Body size or mass is one of the main factors underlying food webs structure. A large number of evolutionary models have shown that indeed, the adaptive evolution of body size (or mass) can give rise to hierarchically organised trophic levels with complex between and within trophic interactions. However, these models generally make strong arbitrary assumptions on how traits evolve, casting doubts on their robustness. In particular, biomass conversion efficiency is always considered independent of the predator and prey size, which contradicts with the literature.  In this paper, we propose a general model encompassing most previous models which allows to show that relaxing arbitrary assumptions gives rise to unrealistic food webs. We then show that considering biomass conversion efficiency dependent on species size is certainly key for food webs adaptive evolution because realistic food webs can evolve, making obsolete the need of arbitrary constraints on traits' evolution. We finally conclude that, on the one hand, ecologists should pay attention to how biomass flows into food webs in models. On the other hand, we question more generally the robustness of evolutionary models for the study of food webs.

\end{abstract}

\section{Introduction}

Predicting the response of
ecological systems to global change through the evolution of
individuals' traits, such as size and feeding rates, is a challenge for modern theoretical ecology \citep{brown2004a,
woodward2005a}. This is especially urgent in the context of fisheries management when
harvesting increases \citep{laugen2014}, or for anticipating how  agrosystems' communities
would evolve in response to agricultural practices change \citep{loeuille2013}.

Roughly, two categories of models have been developed to address these issues and study the
properties of food webs. A first category deals with the stability of a
given food web for a given number of interacting species described by
dynamical systems \citep[\textit{e.g.}][]{brose2006a, otto2007a,
allesina2012, miele2019}. These models often aim at addressing how a
particular interaction affects the stability of a community or the
productivity of an ecosystem. However, in these models,
the species' features are supposed known and fixed. In other words, these models
only consider ecological time scales and neglect evolution and adaptation. In the present paper, we focus on a second category of models which
addresses how and why food web structure and topology emerge on
evolutionary time scales, especially the ones which are variations of
the seminal work by \cite{loeuille2005a}. These models are inspired by
the adaptive dynamics framework \citep{metz1996a, geritz1998a}, {\it
i.e.} mutations affecting one or several traits (especially body size or
mass, or other parameters associated with predation such as predation
preference) are recurrently introduced into the community. The evolution
of traits can lead to evolutionary branching, {\it i.e.} a new species
appears into the community, which affects the whole ecosystem since new
ecological interactions emerge.

Despite the common foundations of these
models  \citep{loeuille2005a}, their assumptions consistently vary
(see Appendix~\ref{sec.review}
for an extended review and Table \ref{table.models}  for a compilation). First, regarding how individuals interact: Cannibalism can be allowed
or not; Predation can be ordered such that bigger species only feed on smaller
ones; In addition to competition for resources, interference competition
can occur or not; And functional responses can follow different functional
forms (Holling Types I or II, Beddington-DeAngelis, etc.). Second, regarding traits' evolution: body size (or mass) can evolve
jointly or not with a trait affecting predation interactions ({\it
e.g.} niche width or prey size difference); Body size (or mass) is
considered on a linear or a logarithmic scale, strongly affecting the
size difference between prey and predators (in particular, assuming a
linear scale limits the minimal size of prey); Traits evolution is
generally constrained by strict boundaries assuming artificial minimal
and maximal possible values; Mutation size
can be large or small, and the distribution of the mutational effects
can differ between models, centred on the parental value or not;
Finally, allometries generally affect
different processes: either birth, death, and/or predation interactions.

In spite of the variability of their assumptions (Table \ref{table.models}), these models typically
show that a food web can emerge with an increase in species number and a
given topology for trophic interactions: a given species can
preferentially consume a subset of the extant species and/or the
resources. These models especially showed that 1) The adaptive
evolution of body size or body mass could be a major mechanism
explaining the structure of food webs, and especially the emergence of
trophic levels \citep[\textit{e.g.}][]{loeuille2005a,brannstrom2011}; 
2) Diversification in trophic networks is promoted by the number of
evolving traits such as predation preference and niche width
\citep{allhoff2013a, allhoff2015a, allhoff2016a, bolchoun2017a} or
abstract traits \citep{ritterskamp2016a}; 3) There can exist a turnover
of species in trophic networks, with species going extinct and replaced
either by new species appeared by mutations, or because of the evolution
of the niche of an extant species \citep{allhoff2015a}.

On the one hand, a shared typical outcome by the different models can be
seen as the evidence that they indeed capture the fundamental mechanisms
responsible for food webs evolution and
diversification. However, on the other hand, because food webs do not
emerge and diversify under similar parameter values, it casts doubts on
their generality and the key mechanisms governing food webs evolution
are yet to be identified. For instance, results are very sensitive
to the size of
mutations considered \citep{allhoff2015a}, and to the choice of
trade-off functions \citep{deMazancourt2004a},
questioning the plausibility and importance of evolutionary branching
as an important process underlying the evolution of trophic networks. 
Furthermore, strong and arbitrary assumptions are shared among all
models  ({\it e.g.} only one trait evolves, or body size is arbitrarily
lower bounded, see Table \ref{table.models}). It is possible that the
properties of these models are due to such arbitrary assumptions, which
would make them much less satisfying and relevant for the study of the
evolution of ecological networks.

Surprisingly, a single feature is common among all models: the
\textit{biomass conversion efficiency} (or conversion factor), {\it
i.e.} the fraction of ingested biomass devoted to the production of
biomass of newborns, is supposed constant. Most
  authors refer to \cite{yodzis1992a} to justify their choice of a constant conversion efficiency, with a value often assumed to be 0.45 for
  herbivores and 0.85 for carnivores. Assuming that the conversion
efficiency is independent of the mass / body size of predators and 
prey is undoubtedly important because it implies that the mass
converted by individuals when consuming prey is increasingly large when
feeding on larger prey, with no limit. This is in contradiction with
empirical data from controlled experiments which show that there is a
trade-off for predators between eating small and large prey
\citep{baras2014a, norin2017a}, suggesting an optimal prey size for which
the conversion efficiency is the highest. This also contradicts the
results of  functional and behavioural ecology models, which predict that the net
energy gain per unit of prey mass is not constant, bounded and can show an optimal values \citep{portalier2018, pawar2019, ho2019}.  This can be
due to a trade-off between the biomass acquired when consuming a prey \textit{vs.} the energy costs of foraging, handling and digesting it, which can depend on the relative size between the predator and the prey (see Fig.~\ref{fig.cartoon}).  \cite{portalier2018} for instance showed that there are prey size limits within which the net energy gain for the predator is positive, which constrains the possible feeding interactions.  It can also be
due to the fact that predators can feed only on a part of a large prey:
the biomass converted from a large prey attains a maximum. How the
efficiency of biomass conversion affects the evolution of food webs has
been ignored by extant models. One can for instance expect that if
predator preference can evolve, it should correspond to the best
compromise between eating small or large prey.

In this paper, our main goal is to explore the importance of the biomass
conversion efficiency in the evolution of food webs. For that purpose,
we propose a unifying framework which encompasses most of the
previously published models. We show that some key properties of the biomass conversion efficiency
govern the evolution of food webs. We then show that relaxing strong assumptions
made by previous models, especially artificial bounds on evolving traits, can give rise to unrealistic trophic networks.
Finally, we discuss the robustness of these models and question the
validity and generality of their predictions, especially if one aims at
using these models for management or conservation purposes.

\section{Model description}
\label{sec:model}

\subsection{A unifying model}
\label{sec:unifying}

Models of food web evolution derived from the seminal work of \cite{loeuille2005a} have a similar structure. For the sake
  of comparison, we propose here a unifying model encompassing most of the models' features in  \cite{loeuille2005a, ingram2009,brannstrom2011,allhoff2013a,allhoff2015a,allhoff2016a,ritterskamp2016a,ritterskamp2016b,bolchoun2017a}. All these
  models assume that new species are added to the food web at discrete mutation times and that between two mutations the densities of
  species follow deterministic dynamics. On a time interval between two mutations where the food web is composed of
  $n$ species, species $i$ is characterized by its average body mass $r_i$ at maturity and its 
population density  $N_i$, for all $i=1\ldots n$. An autotrophic resource (organic or inorganic) is present in the community, {\it i.e.}  a species which does not consume any other species but can be consumed. The  density of the resource is indexed by $i=0$. The basic dynamics for $N_i$ takes
the form
\begin{align}
\label{general.eq}
\frac{\dot N_i}{N_i}
	&=
		\sum_{j = 0}^{n} \lambda_{ij} \, \gamma_{ij}\,N_j
		-\sum_{j=1}^{n}\,\alpha_{ij}\,N_j
		-\sum_{j=1}^{n}\gamma_{ji}\,N_j
		-m_i\,, \qquad i=1,\dots, n
\end{align}
where $\gamma_{ij}$ is the consumption rate of prey $j$ by predator $i$; $m_i$ is the mortality rate of species $i$; $\alpha_{ij}$ is
the interference competition rate between species $i$ and $j$; $\lambda_{ij}$ is the reproduction efficiency due to the consumption of species $j$ by species $i$, {\it i.e.} the per capita reproduction rate of species $i$ per capita of ingested species $j$. This last quantity summarizes biomass conversion and reproduction. The reproduction efficiency
$\lambda_{ij}$ is related to the {\it biomass conversion efficiency} through predation of prey $j$ by predator $i$, denoted
$\xi_{ij}$,
by $\lambda_{ij}=\xi_{ij}\frac{r_j}{r_i}$. The biomass conversion efficiency $\xi_{ij}$ corresponds to the fraction of ingested
  biomass of species $j$ converted through reproduction into biomass of species $i$. It is assumed in particular that $\xi_{ij}<1$ to avoid biomass creation {\it ex nihilo}.

 We denote the density of the autotrophic resource  by $N_0$ and by $r_0$ its (bio-)mass per unit density.
The resource dynamics varies among references,  apparently without significantly affecting the behaviour of the model. In this work, in line with \cite{brannstrom2011}, we consider the following logistic dynamics for the resource density:
\begin{align}
\label{eq.ressource}
\frac{\dot N_0}{N_0} = r_g-k_0\,N_0-\sum_{i=1}^{n}\gamma_{i0}\,N_i\,
\end{align}
where $r_g$ and $k_0$ are the reproductive rate and the intraspecific competition rate of the resource population, respectively.

The ecological parameters $\gamma_{ij}$, $\lambda_{ij}$, $\alpha_{ij}$ and $m_i$ are functions of species' phenotypes characterized by the average body mass $r_i$, the preferred predation distance (in
terms of body mass) $d_i$, and a predation range parameter $s_i$, often called the ``niche width''. It is convenient to express the ecological parameters in terms of the log-body mass $z_i=\ln(r_i/r_0)$ instead of $r_i$, in which case the preferred distance of predation $d_i$ is replaced by its logarithmic expression  $\mu_i = \ln(r_i/(r_i-d_i))$, so that $\gamma_{ij}$ is maximal when $z_j=z_i-\mu_i$, \textit{i.e.} $r_j = e^{z_i-\mu_i} = e^{-\mu_i}\,r_i$.   In other words, the preferred prey biomass of a species is expressed as a fraction of its own biomass. On the logarithmic scale, the corresponding niche width is denoted $\sigma_i$.

These phenotypes can evolve because of recurrent mutations, on the one hand,  and because of the competitive and predation interactions between individuals which induce selection, on the other hand. In particular, $r_i$ (or $z_i$) is considered as an
  evolving trait. Mutations are assumed to be rare, so that Eqs. \eqref{general.eq} and \eqref{eq.ressource} may reach a stationary
state (if it exists) before the next mutation. The ancestral species producing the new (mutant) species is chosen with
probability proportional to their density or biomass. The mutant trait is drawn in a 
distribution which depends on the ancestral trait. The mutant is introduced in the
population at a density very small relatively to the total density of the community.

\subsection{Parameterization: How ecological parameters depend on body mass or size}
\label{sec.model}

For comparison, Table \ref{table.models} compiles assumptions made in models about  how body size (or mass) affect ecological parameters such as predation, competition or reproduction rates.  For reasons detailed in Section~\ref{sec:compar},  we
  consider the following parameterization in our model. For species $i$, the log-body mass $z_i$ and the predation log-distance $\mu_i$ can evolve. We assume
without loss of generality that $r_0=1$, so that $z_i=\log r_i$. The predation rate is given by
\begin{align}
\label{pred.brannstrom}
	\gamma_{ij} = \gamma(z_i-z_j-\mu_i)
		:=
		\frac{\gamma_0}{\sqrt{2\,\pi}\,\sigma_\gamma}\,
		\exp\left(-\frac{(z_i-z_j-\mu_i)^2}{2\,\sigma_\gamma^2}\right)\,.
\end{align} 
In particular, Eq. \eqref{pred.brannstrom} supposes that all species can feed on any other species, but predation of species $j$ by species $i$ is significant only when
  $z_j$ is close to $z_i-\mu_i$, within a range of the order of $\sigma_\gamma$ (Fig.~\ref{fig.pred.rate}). The niche width $\sigma_\gamma$ is assumed fixed.

The interference competition parameter $\alpha_{ij}$ follows a Gaussian function such as
$$
	\alpha_{ij} = \alpha(z_i-z_j)
		:=
		\frac{\alpha_0}{\sqrt{2\,\pi}\,\sigma_\alpha}\,
		\exp\left(-\frac{(z_i-z_j)^2}{2\,\sigma_\alpha^2}\right)\,,
$$ 
where $\sigma_\alpha$ is the competition range on the logarithmic scale.

The mortality rate takes into account allometry, following well-known empirical observations of \cite{peter1983a}, as
$$
	m_i= m(z_i) := m_0\,e^{-0.25\,z_i} = m_0\,r_i^{-0.25}\,.
$$

\begin{figure}
  \includegraphics[width=15cm]{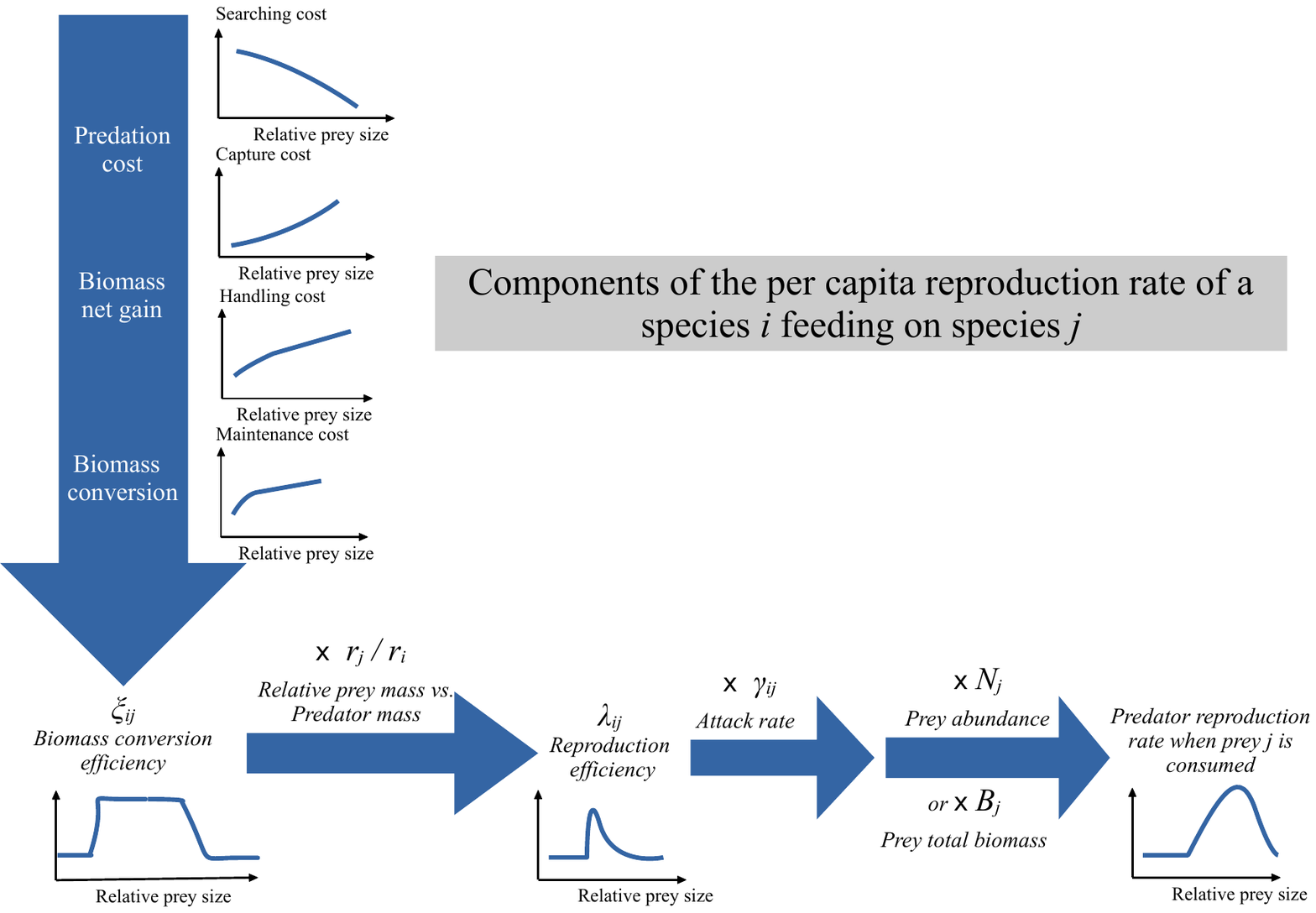}
  \caption{\label{fig.cartoon} Why biomass conversion and reproduction efficiencies depend on both the prey and predators' sizes. Top left: Feeding on a prey needs spending energy for foraging, capturing, handling and then digesting it to acquire biomass. All these ecological processes can depend on the relative size between the prey and the predator as illustrated: the time and energy spent to find a prey can decrease when the prey size increases, while capture and handling costs can increase with prey size (note that the form given to the curve are for illustration only, for more precise functional forms see references in the main text). Bottom left: the biomass conversion efficiency is decomposed into the ecological processes involved in predation interactions. Since these ecological processes vary with prey size, the biomass conversion efficiency should consequently depend on the prey and predators' sizes. The curve for $xi(z_i,z_)$ in particular depicts that there should be a minimum and maximum prey size under and above which biomass conversion efficiency is zero, reflecting that eating a prey that is too small or too big is inefficient because it takes too much energy to forage, capture and handle.  Bottom middle: The form of the reproduction efficiency (per capita) $\lambda_{ij}$ is a direct result of the form of the biomass conversion efficiency $\xi(z_i,z_j)$ weighted by the ratio between the prey and predators' biomass $r_j/r_i$. Bottom right: the predator reproduction rate depends on the reproduction efficiency $\lambda_{ij}$, the attack rate $\gamma_{ij}$, and the prey abundance $N_j$ (or biomass $B_j$).}
\end{figure}
\begin{figure}
\captionsetup[subfigure]{justification=centering}
\begin{center}
\begin{subfigure}{0.45\textwidth}
 \includegraphics[width=7cm]{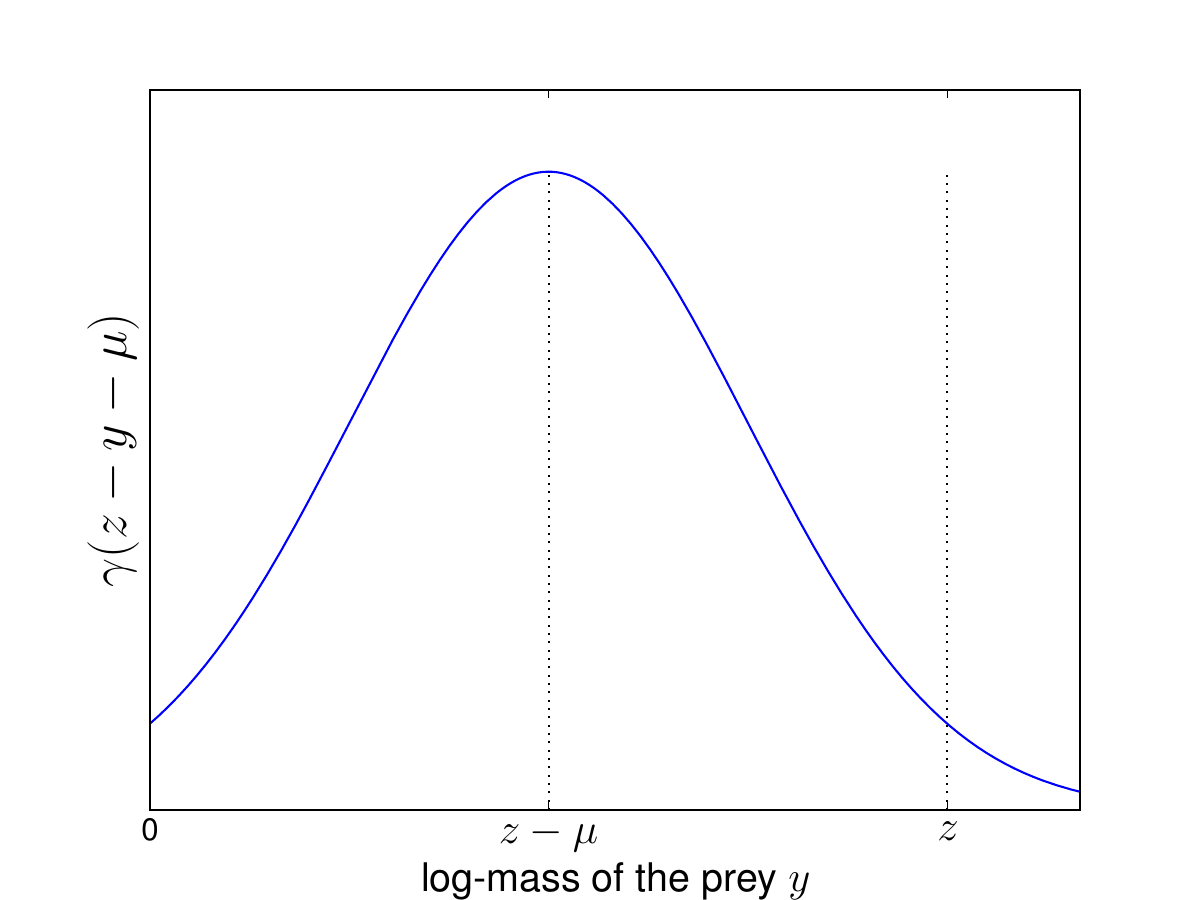}
\caption{\label{{fig.pred.rate.log.scale}}  logarithm scale}
\end{subfigure}
\begin{subfigure}{0.45\textwidth}
 \includegraphics[width=7cm]{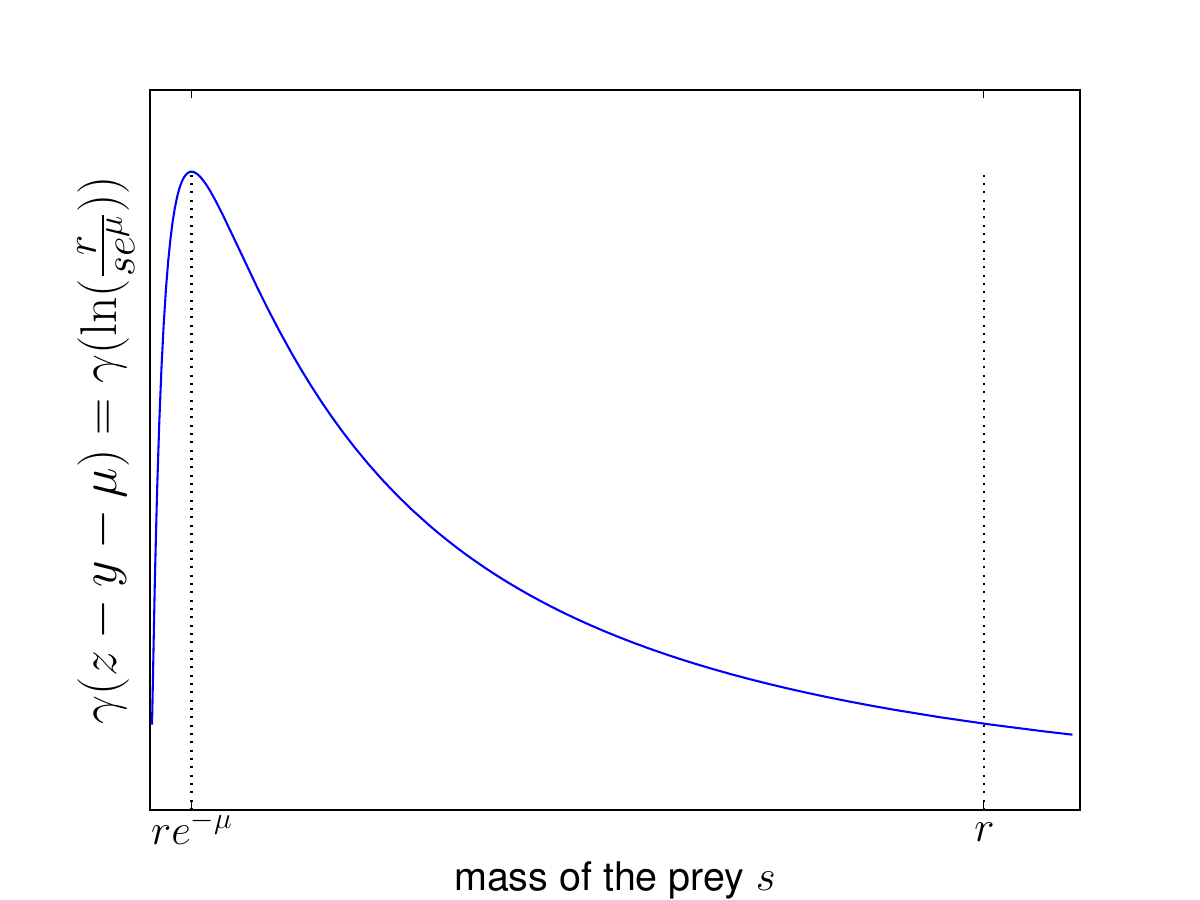}
\caption{\label{{fig.pred.rate.linear.scale}} linear scale}
\end{subfigure}
\end{center}
\caption{\label{fig.pred.rate} Predation rate from Eq. \eqref{pred.brannstrom} represented on the logarithm (left) and linear (right) scales.}
\end{figure}

In all previously published models (Tab. \ref{table.models}), the biomass conversion efficiency $\xi_{ij}$ is assumed to be constant, \textit{i.e.} the biomass
  produced by reproduction is a fixed fraction of the ingested biomass. The reproduction efficiency $\lambda_{ij}$ then increases
with the prey mass / size without limits (see Figure~\ref{fig.ex.xi1}, black thick lines). Experiments rather suggest that there
exist trade-offs between energy gains and loss linked to predation due to foraging, catch, ingestion and digestion
\citep[\textit{e.g.}][]{baras2014a, norin2017a}. Generally, most ecological processes involved in predation interactions depend on both the prey and the predator sizes (see Fig. \ref{fig.cartoon}) as suggested by functional and behavioural ecology models \citep{portalier2018, pawar2019, ho2019}.  In particular, if
individuals feed on larger prey, the hunting and handling costs per unit of biomass is likely to be larger than for smaller prey \citep{portalier2018}. Moreover, large
prey might not be totally ingested by predators. Conversely, if prey are small, a predator has to feed on a large number of
prey and then the handling time (per unit of biomass) becomes critical. This suggests that
$\xi_{ij}$ should depend on the log-masses of the predator $z_i$ and the prey $z_j$, and that  it should  tend to zero when the
relative size $z_i-z_j$ goes to $\pm \infty$ (Fig. \ref{fig.cartoon}). Hence, we assume that the reproduction efficiency is such that
\begin{align}
\label{eq.lambda}
	\lambda_{ij} = \lambda(z_i,z_j) = \frac{e^{z_j}\, \xi(z_i,z_j)}{e^{z_i}}= \xi(z_i,z_j)\,\frac{r_j}{r_i}\,.
\end{align}
In the following, Section~\ref{sec:challenging} is devoted to the analysis  of the general model described above for any functional form of the biomass conversion efficiency $\xi(z_i,z_j)$. 
In  Section~\ref{sec.num.study}, a numerical analysis of the model is performed assuming a specific form of the function $\xi(z_i,z_j)$, depending only on $z_i-z_j$.

\subsection{Comparison with the  parameterization of models derived from \cite{loeuille2005a}}
\label{sec:compar}

In this section, we show how our model encompasses most of the models derived from \cite{loeuille2005a}, focusing on the most important features, excluding the biomass conversion efficiency which was thoroughly discussed previously (see Tab. \ref{table.models} and App. \ref{sec.review} for a full review). Our aim is to minimize biological artificial constraints and arbitrary assumptions. Overall, our model is similar to the one by \cite{brannstrom2011}, except that we consider that both body size and preferred predation distance evolve, and that the biomass conversion efficiency depends on the prey and predators sizes. 

All models, including ours, assume that the main trait driving evolution is body-mass, but there are many differences on other evolving traits.
  It is hard to know which are the most important traits for long term evolution of food webs. We chose to focus on
  log-body mass and predation log-distance because assuming fixed predation distances strongly constrains the dynamics and does not
  allow the evolution of cannibalism, although it is widespread in nature~\citep{fox1975a}. For the sake of simplicity, we kept other ecological
  parameters constant, in particular the niche width $\sigma_\gamma$. Indeed, it has been shown by \cite{ingram2009} and \cite{allhoff2013a}
  that, in models with a non-evolving and abundant resource, niche width generally evolves to very small values unless an {\it ad hoc} trade-off is introduced on niche width, or unless parameters
  are finely tuned.

We do not impose any {\it a priori} constraints about whether or not a given species can feed on another species, contrarily to many models (see columns ``Ordered predation'', ``Cannibalism'' and ``Predation rate'' in Tab. \ref{table.models}): in our model cannibalism is allowed and predation is unordered meaning that predators do not necessarily feed on smaller species. As a consequence, if predation is ordered and if cannibalism is rare in the food web, these are {\it a posteriori} emerging properties derived from the ecological and evolutionary processes.

As shown in Eqs.  \eqref{general.eq} and \eqref{eq.ressource}, we model the dynamics of the density of the species $N_i$. In
\cite{loeuille2005a,allhoff2013a,ritterskamp2016b}, populations dynamics are instead expressed in terms of species' biomass
$B_i=N_i\,r_i$ and assuming that the resource has body-mass $r_0=0$. In Table~\ref{table.models}, the conversion of the dynamics on $B_i$ to a
    dynamics on $N_i$ is done using the assumption that $r_0=0$ and $B_0=N_0$ and the fact that, if for all $i\geq 1$
\begin{align*}
  \frac{\dot B_i}{B_i}
  &=
  \sum_{j = 0}^{n} \lambda^B_{ij} \, \gamma^B_{ij}\,B_j
  -\sum_{j=1}^{n}\,\alpha^B_{ij}\,B_j
  -\sum_{j=1}^{n}\gamma^B_{ji}\,B_j
  -m_i\,,
\end{align*}
then
\begin{equation}
  \label{eq:conversion-size-biomass}
  \lambda^B_{ij} = 
  \begin{cases}
    \frac{r_i}{r_j}\,\lambda_{ij} & \text{ if $j\geq 1$}\\
    r_i \, \lambda_{ij} & \text{ if $j= 0$}
  \end{cases}\,, \qquad
  \gamma^B_{ij} = \frac{\gamma_{ij}}{r_i}\,, \qquad \text{and} \qquad
  \alpha^B_{ij} = \frac{\alpha_{ij}}{r_j}\,.
\end{equation}
Models considering the dynamics of the biomass instead of the density of species hence assume that the size of the autotrophic resource is null. This imposes that the species consuming the autotrophic resource can not be smaller than the resource since size cannot be negative. Here again, if species consuming the autotrophic resource in the food webs are always larger than the resource, it is due to an emerging property of the model.

Several models supposed additional arbitrary constraints on the possible values of evolving
  traits (see column ``Boundaries'' in Tab.~\ref{table.models}). In particular, assumptions about the mutation kernels can strongly constrain the evolution of phenotypes. For example,
  in~\cite{allhoff2015a,allhoff2016a,bolchoun2017a}, both predation distance and niche width can evolve, but mutations only produce
  trait values in a fixed interval. As a consequence, phenotypic traits can be limited to positive values (any mutations introducing
  a negative trait values in the population is arbitrarily discarded). It often appears that these arbitrary boundaries are necessary
  otherwise degenerated trophic webs emerge. For instance, assuming that predation distance can not take negative values
  hinders the evolution of cannibalism or of species feeding on larger prey. In App. \ref{sec.review} we review in more details why such arbitrary boundaries can bias the evolution of food webs.

Finally, in models by \cite{loeuille2005a,ingram2009,allhoff2013a,ritterskamp2016a}, instead of resource dynamics described by Eq. \ref{eq.ressource}, a chemostat equation is assumed for the dynamics of resources, with possible recycling of a fraction $\nu$ of biomass of dead individuals into resources:
\begin{align}
\label{eq.ressource.chemostat}
\dot N_0 = I-eN_0-\sum_{i=1}^{n}\gamma_{i0}N_iN_0
	+\nu\sum_{i=0}^n r_iN_i\left(m_i+ \sum_{j=1}^n \alpha_{ij}N_j
	+ \sum_{j=1}^{i-1}(1-\lambda_{ij})\gamma_{ij}N_j\right),
	\tag{2'}
\end{align}
where $I$ and $e_0N_0$ are the in- and out-flow of resources, respectively. The way the dynamics of the resources is considered does not seem to affect significantly the outcomes of models.

\begin{sidewaystable}
\begin{center}
{\small
\begin{tabular}{|c|c|c|c|c|c|c|c|c|}
\hline
& Evolving phenotypes & Boundaries & Model type & Ordered  predation & Cannibalism & Allometries & Mutations size 
\\
\hline
	$\begin{matrix} 
	\text{LL05,} \\
	\text{RBB16b}\\
	\end{matrix}$
	& mass $r$
	& $0<r$
	& \eqref{general.eq}-\eqref{eq.ressource.chemostat}
	& yes
	& no
	& birth \& death
	& large
\\
\hline
	IHS09
	& $\begin{matrix} 
	\text{mass } r\\
	\text{niche width } s
	\end{matrix}$
	& $\begin{matrix} 
		0<r\\
		0< s	
	   \end{matrix}$
	& $\begin{matrix}
		\eqref{general.eq}-\eqref{eq.ressource.chemostat}\\
		\text{with ST I-II}\\
	  \end{matrix}$
	& yes
	&  no
	& birth \& death
	&large
\\
\hline
	BLLD11
	& log-mass $z$
	& $z \in \mathbb{R}$
	& \eqref{general.eq}-\eqref{eq.ressource}
	& no
	& yes
	& death
	& small
\\
\hline
	AD13
	& $\begin{matrix} 
		\text{mass } r\\
		\text{predation distance } d\\
		\text{niche width } s	
	   \end{matrix}$
	& $\begin{matrix} 
	0<r\\
	0<d\\
	0<s
	\end{matrix}$
	& \eqref{general.eq}-\eqref{eq.ressource.chemostat}
	& yes
	& no
	& birth \& death
	& large
\\
\hline
	$\begin{matrix} 
	\text{ARRDG15,} \\
	\text{AD16}\\
	\text{\& BDA17}
	\end{matrix}$
	& $\begin{matrix} 
		\text{log-mass } z\\
		\text{predation log-distance }  \mu\\
		\text{niche width } \sigma	
	   \end{matrix}$
	& $\begin{matrix} 
	z \in \mathbb{R}\\
	0.5<\mu<3 \\
	0.5<\sigma<1.5 
	\end{matrix}$
	& $\begin{matrix}
		\eqref{general.eq}-\eqref{eq.ressource}\\
		\text{with}\\
		\text{BDA}
	  \end{matrix}$	
	& no
	& yes
	& predation \& death
	& large
\\
\hline
	RBB16a
	& $\begin{matrix} 
	\text{log-mass } z\\
	\text{abstract trait } x
	\end{matrix}$
	& $z \in \mathbb{R}$
	& $\begin{matrix} 
	\eqref{general.eq}-\eqref{eq.ressource.chemostat}\\
	  \text{with }\nu=0
	  \end{matrix}$
	& no
	& no 
	& predation \& death
	& large
\\
\hline
	$\begin{matrix}
	\text{Our}\\
	\text{model}
	\end{matrix}$
	& $\begin{matrix} 
		\text{log- mass } z\\
		\text{predation log-distance } \mu\\
	   \end{matrix}$
	& $z,\mu \in \mathbb{R}$
	& \eqref{general.eq}-\eqref{eq.ressource}
	& no
	& yes
	& death
	& small
\\
\hline
\end{tabular}
}

\bigskip

{\small
\begin{tabular}{|c|c|c|c|c|c|c|c|c|}
\hline
& Predation rate $\gamma_{ij}$ 
& Production efficiency $\lambda_{ij}$ 
& Competition rate $\alpha_{ij}$ 
& Death rate $m_i$
& Mutations
\\
\hline
	$\begin{matrix} 
	\text{LL05,} \\
	\text{RBB16b}\\
	\end{matrix}$
	& $\begin{matrix} 
	0 \text{ if $r_i\leq r_j$, otherwise:}\\
	r_i\,\gamma_0\,G_{s}(r_i-r_j-d)
	\end{matrix}$
	& $\begin{cases} 
          \lambda_0\,r_i^{-0.25}\,\frac{r_j}{r_i} & \text{if $j\neq 0$}\\
	\lambda_0\,r_i^{-1.25}  & \text{if $j=0$}
	\end{cases}$		
	& $\begin{cases} 
	\alpha_0\,r_j & \text{if $|r_i-r_j|\leq \beta$}\\
	0 & \text{otherwise,}
	\end{cases}$
	& $m_0\,r_i^{-0.25}$
	& $r'\sim \mathcal{U}([0.8\,r;1.2\,r])$
\\
\hline
	IHS09
	& $\begin{matrix} 
	0 \text{ if $r_i\leq r_j$, otherwise:}\\
	r_i\,\gamma_0(s_i)\,G_{s_i}(r_i-r_j-d)
	\end{matrix}$
	& $\begin{cases} 
          \lambda_0\,r_i^{-0.25}\,\frac{r_j}{r_i} & \text{if $j\neq 0$}\\
	\lambda_0\,r_i^{-1.25}  & \text{if $j=0$}
	\end{cases}$
	& $\alpha_0\,r_j\,G_{\sigma_\alpha}(r_i-r_j)$
	& $m_0\,r_i^{-0.25}$
	& $\begin{matrix} 
	r'\sim G_{\sigma_r}(r'-r)\\
	s'\sim G_{\sigma_s}(s'-s)
	\end{matrix}$
\\
\hline
	BLLD11
	& $\gamma_0\,G_{\sigma_{\gamma}}(z_i-z_j-\mu)$
	& $\lambda_0\,e^{z_j-z_i}$
	& $\alpha_0\,G_{\sigma_\alpha}(z_i-z_j)$
	& $m_0\,e^{-0.25\,z_i}$
	& $z'\sim G_{0.01}(z'-z)$
\\
\hline
	AD13
	& $\begin{matrix} 
	0 \text{ if $r_i\leq r_j$, otherwise:}\\
	r_i\,\gamma_0\,G_{s_i}(r_i-r_j-d_i)
	\end{matrix}$
	& $\begin{cases} 
	\lambda_0\,r_i^{-0.25}\,\frac{r_j}{r_i} & \text{if $j\neq 0$}\\
	\lambda_0\,r_i^{-1.25} & \text{if $j=0$}
	\end{cases}$		
	& $\begin{cases} 
	\alpha_0\,r_j & \text{if $|r_i-r_j|\leq \beta\,\max\{r_i,r_j\}$}\\
	0 & \text{otherwise,}
	\end{cases}$
	& $m_0\,r_i^{-0.25}$
	& $\begin{matrix} 
	r'\sim \mathcal{U}([0.8\,r;1.2\,r])\\
	d'\sim \mathcal{U}([0.9\,d;1.1\,d])\\
	s'\sim \mathcal{U}([0.9\,s;1.1\,s])
	\end{matrix}$
\\
\hline
	$\begin{matrix} 
	\text{ARRDG15,} \\
	\text{AD16,}\\
	\text{\& BDA17}
	\end{matrix}$
	& $\gamma_0\,e^{0.75\,z_i}\,G_{\sigma_i}(z_i-z_j-\mu_i)$
	& $\lambda_0\,e^{z_j-z_i}$
	& $\begin{matrix} 
	\alpha_0+\alpha'_0 \text{ if $i=j$, otherwise:}\\
	\alpha_0\,\sigma_i\,G_{\sqrt{\sigma_i^2+\sigma_j^2}}(z_i-\mu_i-z_j+\mu_j)
	\end{matrix}$	
	& $m_0\,e^{-0.25\,z_i}$
	& $\begin{matrix} 
	z'\sim \mathcal{U}([z-0.7;z+0.7])\\
	\mu'\sim \mathcal{U}([0.5;3])\\
	\sigma'\sim \mathcal{U}([0.5;1.5])
	\end{matrix}$
\\
\hline
	RBB16
	& $\begin{matrix} 
	0 \text{ if $i=j$, otherwise:}\\
	\gamma_0(x_i,x_j)\,e^{0.75\,z_i}\,G_{\sigma}(z_i-z_j-\mu)
	\end{matrix}$
	& $\lambda_0\,e^{z_j-z_i}$
	& $\alpha_0(x_i,x_j)\,e^{z_j}\,G_{\sqrt{2}\,\sigma}(z_i-z_j)$
	& $m_0\,e^{-0.25\,z_i}$
	& $z'\sim \mathcal{U}([z-0.7;z+0.7])$
\\
\hline
	$\begin{matrix}
	\text{Our}\\
	\text{model}
	\end{matrix}$
	& $\gamma_0\,G_{\sigma_\gamma}(z_i-z_j-\mu)$
	& $\xi(z_i-z_j)\,e^{z_j-z_i}$
	& $\alpha_0\,G_{\sigma_\alpha}(z_i-z_j)$
	& $m_0\,e^{-0.25\,z_i}$
	& $\begin{matrix} 
	z'\sim G_{\sigma_z}(z'-z)\\
	\mu'\sim G_{\sigma_\mu}(\mu'-\mu)
	\end{matrix}$
\\
\hline
\end{tabular}
}
\end{center}
\caption{Food web models and predation, production efficiency (biomass conversion + reproduction), competition, death and mutation functions in models of \cite{loeuille2005a} (LL05), \cite{ritterskamp2016b} (RBB16b), \cite{ingram2009} (IHS09), \cite{brannstrom2011} (BLLD11), \cite{allhoff2013a} (AD13), \cite{allhoff2015a} (ARRDG15), \cite{allhoff2016a} (AD16), \cite{bolchoun2017a} (BDA17), \cite{ritterskamp2016a} (RBB16a) and our model.
BDA means Beddington-DeAngelis functional response, ST I-II means Saturating Type I or II functional response,
$G_\sigma(x)=\frac{1}{\sqrt{2\,\pi}\,\sigma}\,
		\exp\left(-\frac{x^2}{2\,\sigma^2}\right)$ is the centred Gaussian density with variance $\sigma^2$ and
                $\mathcal{U}([a,b])$ is a uniform distribution on the interval $[a,b]$. When models are expressed in terms of
                  biomass, we convert the different parameters using~\eqref{eq:conversion-size-biomass}. }
\label{table.models}
\end{sidewaystable}

\subsection{Criteria for deciding whether a food web is realistic or not}
\label{sec:relevant}

The assumptions made about which traits evolve, the  boundaries of the phenotypic space, and the form of the   biomass conversion efficiency strongly affect the evolutionary dynamics of food webs. This highlights the
  difficulty to define what one should consider as a realistic food web.
   Of course, a naive answer is that we expect food
webs emerging from the models to look like ``real food webs''. However, how it translates into required qualitative and quantitative
features are rarely precisely defined or justified. Generally, authors are satisfied when two conditions are fulfilled:
diversification occurs such that many species accumulate, and species feed on each other. What are the required topologies is, however,
rarely thoroughly discussed. For instance, \cite{loeuille2005a} considered as satisfying food webs showing a hierarchical linear
structure with larger species feeding on smaller ones, even though there is no such simple food web in natural ecosystems.
\cite{allhoff2015a} compare various topological features of simulated and empirical food webs, but it is hard to know which
  particular topological properties are needed to characterise realistic food webs.

Without a generally admitted definition of what is a realistic food web or not, we proceeded the other way around by clearly defining
which food webs emerging from our model will be considered as unrealistic: when 1) all species evolve to a size smaller than the autotrophic resources (hereafter called a {\it negative food web}), or 2) when species mostly feed on the autotrophic resource but little on the other
species (hereafter called a {\it trivial food web}). In a trivial food web, predation interactions can be totally absent,
  or the steady state community can be composed of a single species feeding on the resource. Hence, we consider that a food web is
 realistic when species remain larger than the resource and when at least one species mostly feed on
other species. These simple rules allow us to limit {\it a priori} assumptions about the expected
resulting food webs. In sections \ref{sec:challenging} and \ref{sec.num.study}, we show that some properties of biomass conversion efficiency are key for the evolution of realistic food webs even without imposing arbitrary constraints.

\section{Why biomass conversion efficiency is key to avoid the evolution of unrealistic food webs}
\label{sec:challenging}

Our goal in this section is to identify general conditions on the biomass conversion efficiency $\xi_{ij}=\xi(z_i,z_j)$ and reproduction efficiency
  $\lambda_{ij}=\lambda(z_i,z_j)$ (Section~\ref{sec.model}) for the evolution of realistic food webs, or more precisely for preventing the emergence of negative or trivial food webs.

\subsection{Negative food webs evolve when the biomass conversion efficiency  weakly varies with prey size}
\label{sec.a=1}

The invasion fitness of a mutant $(y,\eta)$ in the food web $(z_i,\mu_i)_{1\leq i \leq n}$ is given by
\begin{align}
f(y,\eta)
\nonumber
	&=
		\sum_{i=0}^{n} \lambda(y,z_i)\,\gamma(y-z_i-\eta)\,N_i^*
\\
\label{eq.f}
	&\quad
		-\sum_{i=1}^{n}\gamma(z_i-y-\mu_i)\,N_i^*
		-\sum_{i=1}^{n} \alpha(z_i-y)\,N_i^* - m(y)
\end{align}
where $N_0^*$ and $(N_i^*)_{1\leq i \leq n}$ are respectively the resource concentration and the population densities at the
stationary state of the food web, {\it i.e.}\  such that Eq.~\eqref{general.eq} and \eqref{eq.ressource} vanish. The sign of the invasion fitness
determines whether a species $(y,\eta)$ can invade the food web $(z_i,\mu_i)_{1\leq i \leq n}$ or not.
In particular we have the classical relation $f(z_i,\mu_i)=0$ for any $i\in \{ 1,\dots,n \}$. 

The analysis we develop here follows the adaptive dynamics framework \citep{metz1996a, geritz1998a, dieckmann1996a, champagnat2011a,
  champagnat2001a}. Under the assumption that mutations are small, the direction of evolution of a given species $(z,\mu)\in (z_i,\mu_i)_{1\leq i \leq n}$ is
governed by the fitness gradient $\nabla f(y,\eta)|_{(y,\eta)=(z,\mu)}$, and the properties of evolutionary equilibria,
such as evolutionary branching, are governed by the second order derivatives $\partial^2_yf$ and $\partial^2_\eta f$.

\medskip

In cases where the predation preference trait $\mu$ evolves  sufficiently fast,
the evolution of the food web is mostly governed by the partial derivative of the conversion efficiency
$$
\partial_y \lambda(y,z)= \left[ \frac{\partial_y\xi(y,z)}{\xi(y,z)}-1\right] \lambda(y,z)\,.
$$
In this case, the fitness gradient can be approximated by (see Appendix~\ref{app.slope.xi} for details)
\begin{align}
\label{eq.approx.partialf}
\partial_1 f(z,\mu)=\partial_y f(y,\eta)|_{(y,\eta)=(z,\mu)}\approx\partial_1\lambda(z,\tilde z)\,\gamma(0)\,\tilde N^* - m'(z),
\end{align}
where $\tilde z$ and $\tilde N^*$ respectively are the size of the major prey of species $(z,\mu)$,
and the total density of the major prey of  species $(z,\mu)$ (possibly including the autotrophic
resource). If $\tilde N^*$ is large enough, the last quantity has the same sign as $\partial_1\lambda(z,\tilde z)$.

 All models compiled in Tab.~\ref{table.models} assume $\xi$ is constant
so that $\partial_1\lambda(z,\tilde{z})=-\lambda(z,\tilde{z})$, which implies
that decreasing predator's body mass while the prey size remains constant necessarily yields a higher reproductive output and a gain
in fitness. Simultaneously, predator's feeding preference will evolve to larger prey's body mass.
This is a consequence of the absence of trade-off on the biomass conversion efficiency, which leads to a strong benefit on the
reproductive efficiency $\lambda$ when feeding on large species (see Figure~\ref{fig.ex.xi1}).

For a negative food web to emerge, a species $(z,\mu)$ close to the point $(0,0)$ (hereafter called a {\it resource-like} species) needs to cross the point $(0,0)$.
    When this occurs, the autotrophic resource and the resource-like species can be considered as a single species for which the
competition from the other species is negligible if other species have log-size sufficiently far from 0. Under these assumptions, the fitness gradient experienced by
  the resource-like species is approximated by
\begin{align*}
\partial_1 f(0,0)
	&\approx
		 \partial_1\lambda(0,0)\,\gamma(0)\,(N_0^*+N^*_1)  - m'(0)
\end{align*}
where $N^*_1$ is the density of the resource-like species. Hence, provided that $m'(0)$ is small enough, the sign of $\partial_1 f(0,0)$ is mainly given by the sign of $\partial_1\lambda(0,0)$.
Therefore, the resource-like species will cross $(0,0)$, initiating the emergence of a negative food web, when
  $\partial_1\lambda(0,0)\lessapprox 0$. Expressed in terms of biomass conversion efficiency, this corresponds to the condition
\[
\frac{\partial_1\xi(0,0)}{\xi(0,0)}\lessapprox 1.
\]
In other words, if the biomass conversion efficiency increases too slowly with the increase of the prey size, our model predicts that species smaller than the autotrophic species should evolve, because consuming a larger prey is always favored for a smaller predator. This finally results in the emergence of a negative food web. 

This result is illustrated in Section~\ref{subsec.constant.conv.eff} (see also Appendix~\ref{sec.relaxConstraints}), where species evolve toward very small body mass and the
  stationary state of food webs shows negative structures. As shown in Table~\ref{table.models}, in all models
    where the predation (log-)distance can evolve, this unrealistic behaviour is not observed because of artificial constraints on
    the trait values or mutation kernels.

\subsection{Trivial food webs emerge when  biomass conversion efficiency is large}
\label{sec.sensitivity.b}

    We now investigate conditions on the biomass conversion efficiency which favour the emergence of trivial food webs, composed of a
    single species. This means that, when only two coexisting species remained in the food web, one of them was competitively
    excluded by the other. Hence we consider two species $(z,\mu)$ and $(y, \eta)$ and apply the competitive exclusion
  principle: species
  $(z,\mu)$ is excluded by species $(y,\eta)$ if the invasion fitness $f(z,\mu)$ of species $(z,\mu)$
  in the food web containing only species $(y,\eta)$ is negative, where
\begin{align}
\label{eq.f.two.species}
  f(z,\mu)&=\lambda(z,0)\gamma(z-\mu)N^*_0 + \lambda(z,y)\gamma(z-y-\mu)N^*_1
 \nonumber \\
  	&\quad -\alpha(y-z)N^*_1-\gamma(y-z-\eta)N^*_1-m(z),  
\end{align}
and $N^*_0$ and $N^*_1$ are the equilibrium densities of resource and species $(y,\eta)$ respectively, when
species $(z,\mu)$ is extinct. For $\lambda(y,0)$ not too small (so that $N_1^*>0$), 
\[
N^*_1=\frac{\lambda(y,0)\gamma(y-\eta)\frac{r_g}{k_0}-m(y)}{\frac{\lambda(y,0)\gamma(y-\eta)^2}{k_0}+(1-\lambda(y,y))\gamma(-\eta)+\alpha(0)}
\quad\text{and}
\quad
N^*_0 =\frac{r_g-\gamma(y-\eta)N^*_1}{k_0}.
\] 

The sign of~\eqref{eq.f.two.species} is hard to evaluate without further information on the values of $(z,\mu)$ and $(y,\eta)$.
  We observe in Section~\ref{sec.simu.impact.b} that the typical situation where the food web evolves to a trivial one corresponds to
  the case where species $(y,\eta)$ is a resource-like species, that is $(y,\eta)\approx (0,0)$. In this case, the sign of
$f(z,\mu)$ is approximately given by a criterion on the ratio
  $\lambda(y,0)/\lambda(z,0)\approx\lambda(0,0)/\lambda(z,0)$ (see Appendix~\ref{app.b.xi} for computation details). More precisely,
the fitness is negative, which means that species $(z,\mu)$ is excluded by the resource-like species
if
\begin{equation}
\label{extinction.criteria.rls}
	 \frac{\lambda(0,0)}{\lambda(z,0)} \geq 
		\frac{(\gamma(0)+\alpha(0))N^*_1+m_0}{(\alpha(-z)+\gamma(-z))\, N^*_1+m(z)}
		\frac{\gamma(z-\mu)}{\gamma(0)}\, ,
\end{equation}
where $N^*_0$ and $N^*_1$ are the equilibrium densities of the autotrophic resource and the resource-like species respectively,
when species $(z,\mu)$ is extinct. The criterion~\eqref{extinction.criteria.rls} can be expressed in terms of biomass conversion efficiency as follows:
\begin{equation}
\label{extinction.criteria.xi.rls}
	 \frac{\xi(0,0)}{\xi(z,0)} \geq 
		\frac{(\gamma(0)+\alpha(0))N^*_1+m_0}{(\alpha(-z)+\gamma(-z))\, N^*_1+m(z)}
		\frac{\gamma(z-\mu)}{\gamma(0)}\,e^{-z}.
\end{equation}

The right-hand-side of~\eqref{extinction.criteria.rls} still depends on $z$ and $\mu$. However, we will observe in the
  simulations of Section~\ref{sec.simu.impact.b} that this quantity stays within a limited range when varying $\lambda(0,0)$. This shows
  that a threshold exists on the quantity $\frac{\lambda(0,0)}{\lambda(z,0)}$: if it is too large, competitive
  exclusion of species $(z,\mu)$ occurs, leading to a trivial food web where only remains a single species feeding on the autotrophic resource.

\section{Numerical study}
\label{sec.num.study}

\begin{figure}
\begin{center}
 \includegraphics[width=7cm]{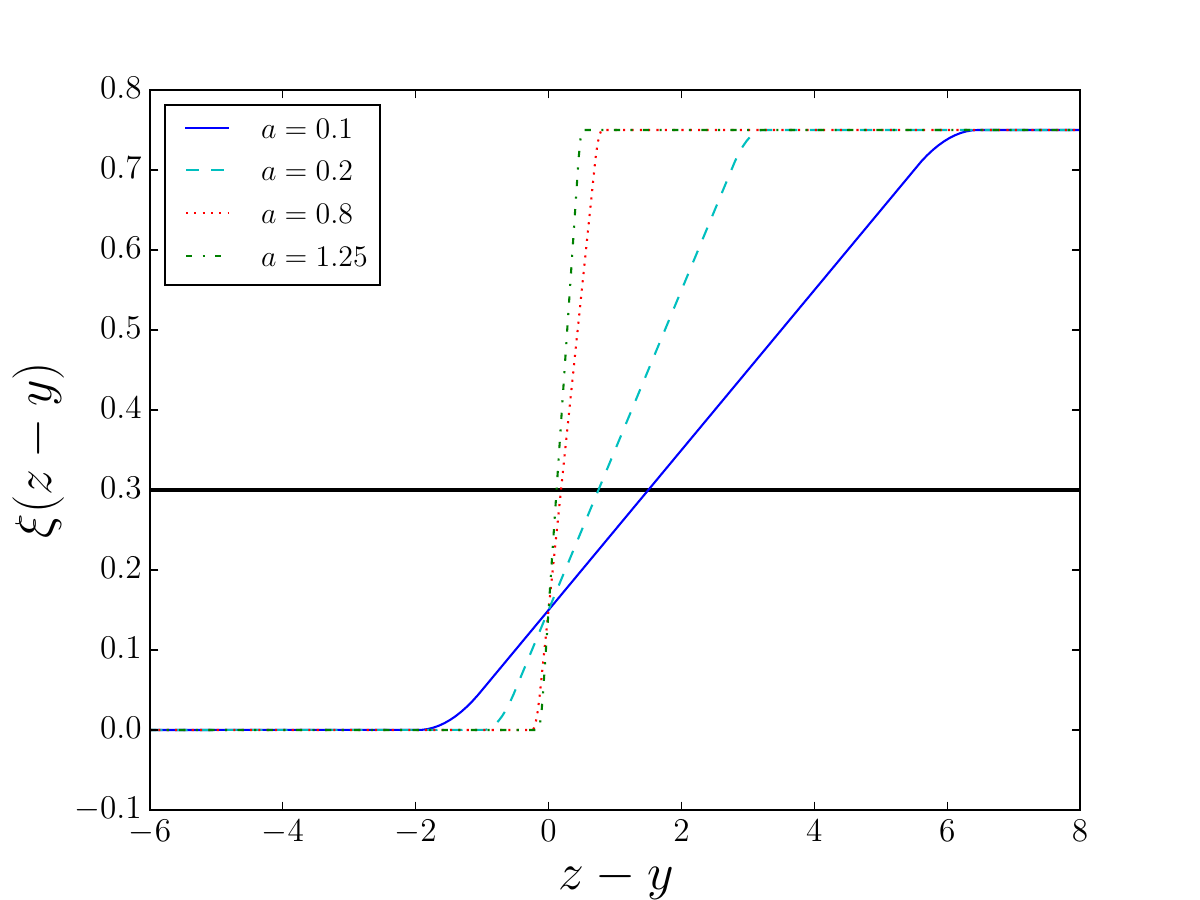}
 \includegraphics[width=7cm]{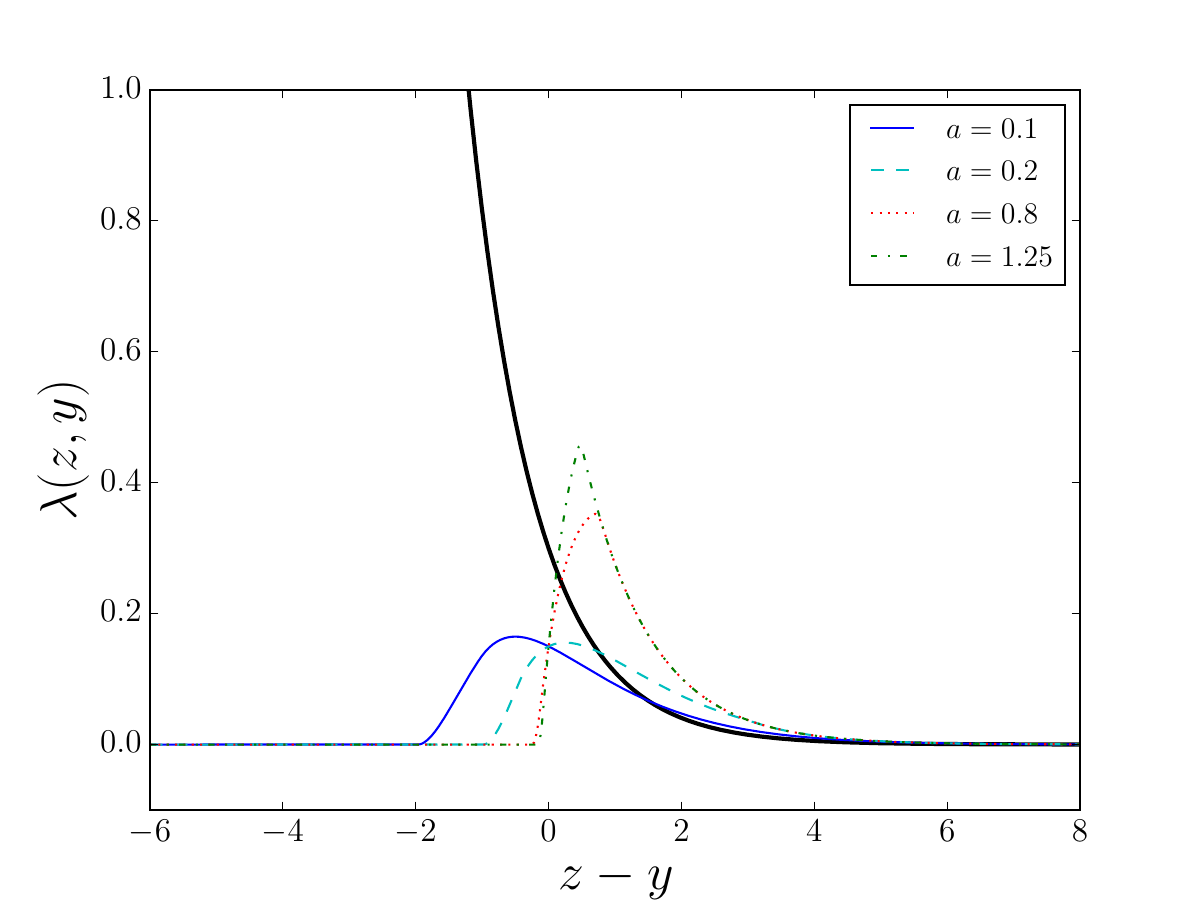}
\end{center}
\caption{\label{fig.ex.xi1}Possible shapes of the trade-off in the biomass conversion efficiency $\xi$ (left) and the reproduction efficiency $\lambda$ (right) defined by Eq.~\eqref{ex.xi1} and
  \eqref{eq.lambda} respectively, with $\xi_{\max}=0.75$ and $b=0.15$ and for the
  model of \cite{brannstrom2011}, that is $a=0$ and $b=0.3$ (black thick line).}
\end{figure}

In this section we will illustrate with numerical studies the general results obtained in the previous section about the properties of the biomass conversion for which realistic food webs can evolve. We consider the following family of functions $\xi$, parameterised by $a,b >0$ and $\xi_{\max}\in(0,1)$:
\begin{align}
\label{ex.xi1}
\xi(z,y) = \xi(z-y) & =
	\begin{cases} 
	0 & \text{if $z-y \leq -1/a$,}\\
	a\,b\,(z-y) + b & \text{if $-1/a \leq z-y \leq \frac{\xi_{\max}-b}{a\,b}$,}\\
	\xi_{\max} & \text{if $z-y \geq \frac{\xi_{\max}-b}{a\,b}$.}
	\end{cases}
\end{align}
$\xi$ is a linear function with slope $a\,b$, such that $\xi(0)=b$ and truncated below $0$ and above $\xi_{\max}$, with
$b<\xi_{\max}$. This last condition means that predation becomes harder in terms of conversion efficiency when feeding on 
  larger prey in an interval of prey sizes containing the predator size.
Examples of such functions are given in Figure~\ref{fig.ex.xi1}.
To avoid problems of irregularity of fitness functions, simulations shown in Figures~\ref{fig.xi.lineaire}-\ref{fig.b} are realised with a regularised version of the previous curves
(see Appendix~\ref{app.reg.xi}).

We assume that mutations occur at regular time units; the species $(z,\mu)$ producing a mutant is drawn proportionally to its individual density (number of individuals); the mutant $(z',\mu')$ is drawn such that $z'$ and $\mu'$ are independent and Gaussian with means $z$ and $\mu$ and variances $\sigma_z^2$ and $\sigma_\mu^2$ respectively. 
The resulting food webs are represented with  an edge  drawn between predator $i$ and 
prey $j$ (or a loop if $j=i$) if predation of $j$ by
$i$ (or cannibalism if $j=i$) is responsible for more than 10\% (5\% for dashed edges) of the reproduction of species $i$.
On Figures~\ref{fig.branstromSC.evol_mu} to \ref{fig.b}, a green edge means that the bigger species feeds on the smaller one, and conversely for magenta edges. Vertical edges represent the resource consumption. More details about simulations methods are given in App.~\ref{app.simu.method}.

\medskip

Our numerical study will focus on the influence on the
  evolutionary dynamics of food webs of the parameters  $a$, $b$, $\xi_\text{max}$, $\sigma_z$ and $\sigma_\mu$ which determine the functional form of the biomass conversion efficiency. All other parameters are the same as in \cite{brannstrom2011}, except for the range of competition $\sigma_\alpha$ which is a bit smaller
in order to favour diversification events in the food web evolution (see Appendix~\ref{app.competition}).
These parameters are given in Table~\ref{table.parametres}. 

\begin{table}[h]
\begin{center}
\begin{tabular}{|cl|c|}
	\hline
    Parameters  & & Values \\
    \hline
    $\sigma_\gamma$ & width of predation kernel  			&	1.5 \\
    $\gamma_0$	 & amplitude of predation kernel 		&	10 \\
    $\sigma_\alpha$&  width of competition kernel  		&	0.5\\
   $\alpha_0$  & amplitude of competition kernel 				&	1\\
	 $m_0$ & proportionality constant of death rate 					&	0.1\\
     $r_g$	 & reproductive rate of  resource			&	10 \\
     $k_0$ &  intraspecific competition rate of  resource 				&	0.01\\
    \hline
\end{tabular}
\end{center}
\caption{Simulation parameters.}
\label{table.parametres}
\end{table}

\subsection{Constant conversion efficiency: emergence of negative food webs}
\label{subsec.constant.conv.eff}

We illustrate the emergence of unrealistic food webs  when the conversion efficiency $\xi$ is constant. In this case, our model corresponds to the model of
  \cite{brannstrom2011} but where both $z$ and $\mu$ can evolve. Figure~\ref{fig.branstromSC.evol_mu} shows
simulations with much smaller mutations on $\mu$ than $z$. The food web initially
evolves as observed by \cite{brannstrom2011} without mutations on $\mu$: several evolutionary
branchings occur and the food web gets structured. However, the smallest species progressively evolves to smaller body size and predation 
  log-distance until they both become negative. This means that this species feeds on a larger prey: the resource. After this, the
richness of the positive part of the food web, composed of the species with
positive body masses and positive predation log-distance, becomes smaller and the negative part of the food web progressively diversifies, producing a negative food web with
more and more species. The stationary state of the food web is not reached at the end of the
simulation, as the food web seems to evolve similarly endlessly. As discussed below, the evolution to a negative food web is due to the fact that parameter
  $a$ is 0 in this case. As mentioned in Section~\ref{sec:compar}, this problem did not occur in
    other models allowing evolution of both traits $z$ and $\mu$ because of
  artificial constraints on the range of phenotypes accessible by mutations (see also Appendix~\ref{sec.relaxConstraints}).

\setlength{\unitlength}{1cm}

\begin{figure}
\begin{center}
\begin{subfigure}{0.32\textwidth}
\begin{picture}(0,3.82)
 \put(0,0){\includegraphics[width=5.1cm]{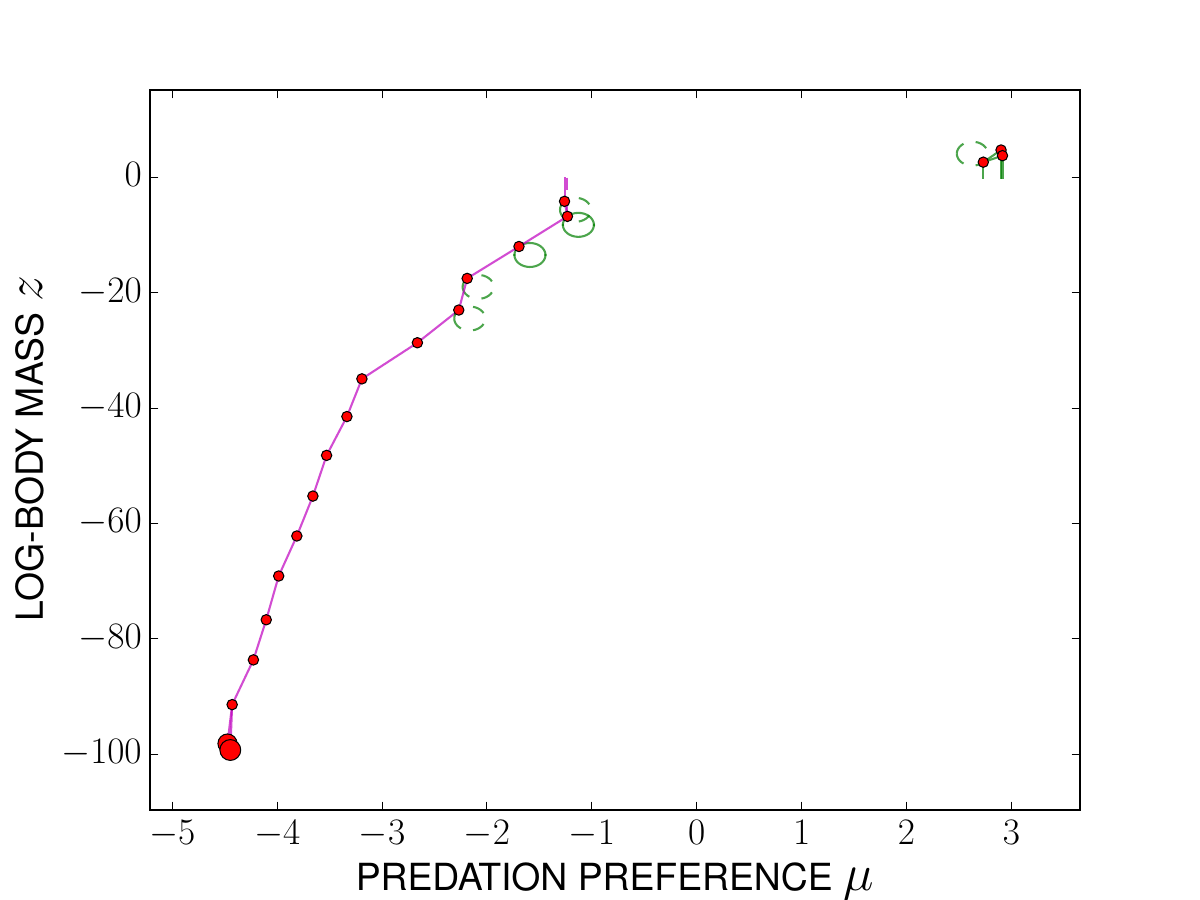}}
 \put(1.8,0.5){\includegraphics[width=2.5cm]{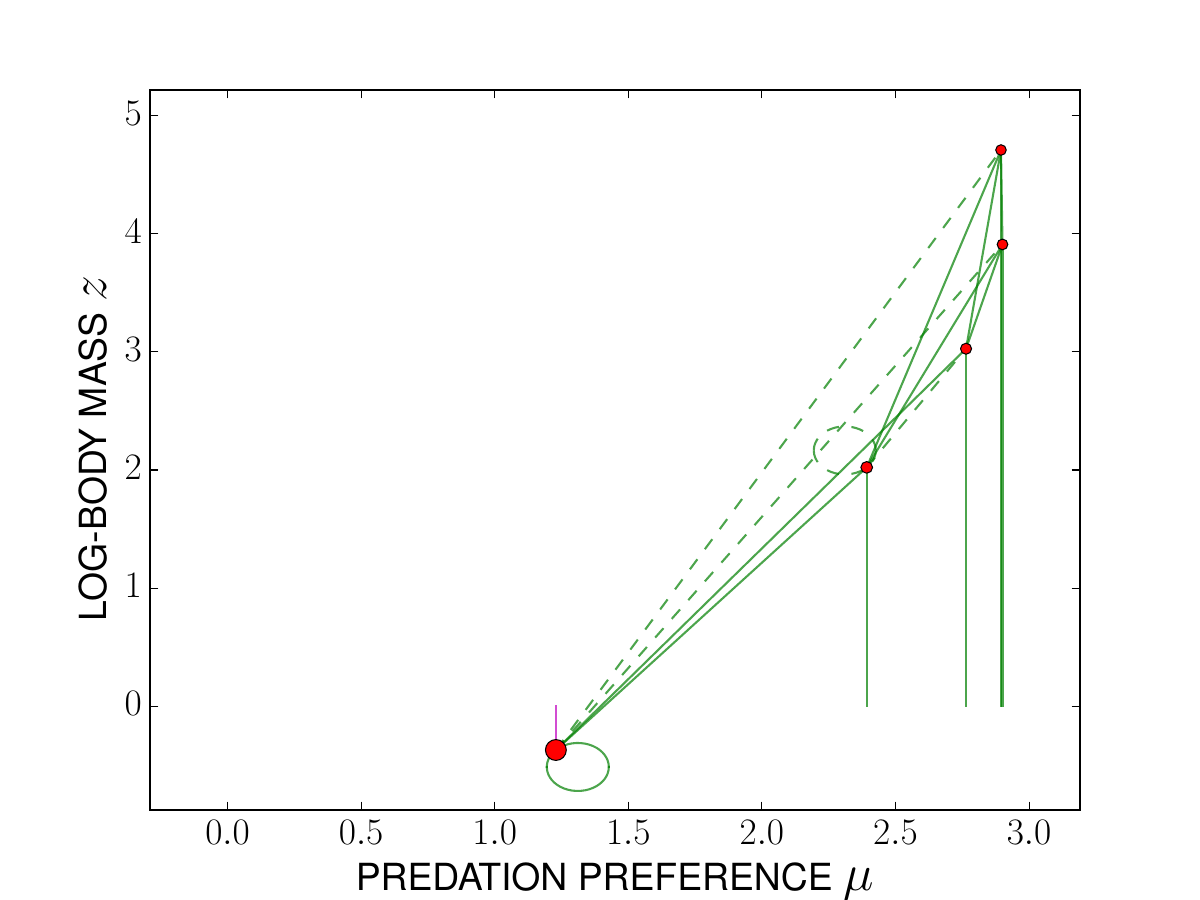}}
\end{picture}
\end{subfigure}
\begin{subfigure}{0.32\textwidth}
\includegraphics[width=5.1cm]{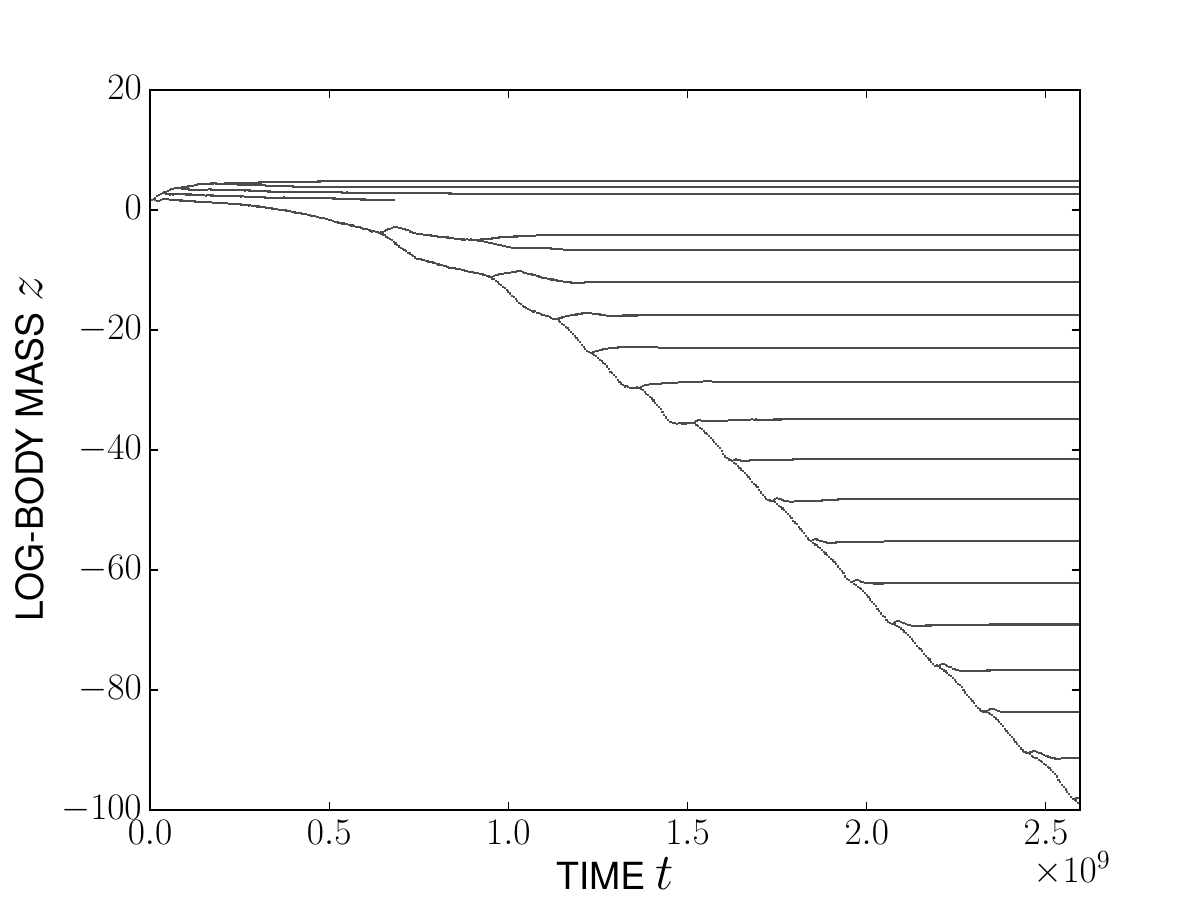}
\end{subfigure}
\begin{subfigure}{0.32\textwidth}
\includegraphics[width=5.1cm]{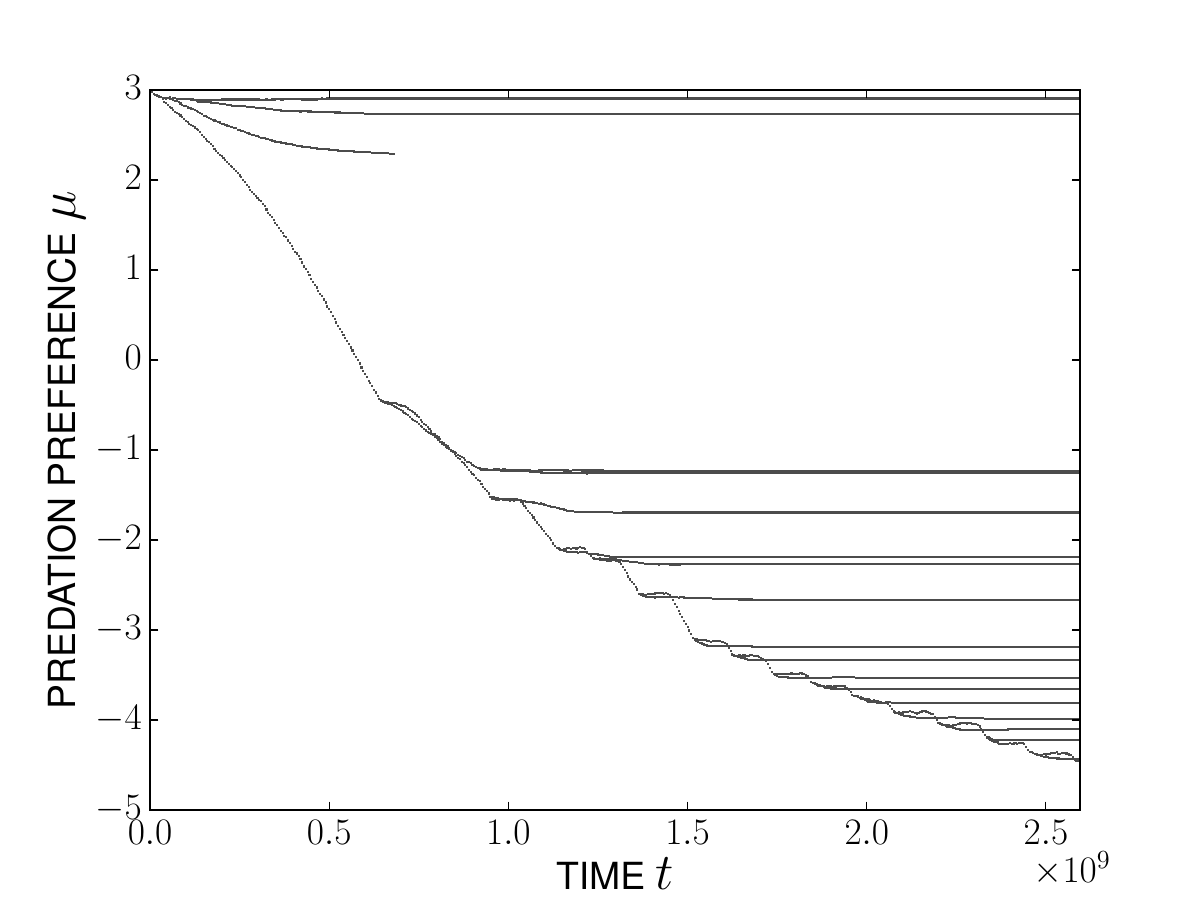}
\end{subfigure}
\end{center}
\caption{\label{fig.branstromSC.evol_mu} Food web evolution with constant biomass conversion efficiency. From left to
  right: food web at final time of the simulation ($T=2.6\,10^9$, overlay at intermediate time $t=4.\,10^8$); evolution of the log-mass $z$; evolution of the
  predation preference $\mu$,  letting $z$ and $\mu$ evolving in the model of \cite{brannstrom2011} (i.e. model of Section~\ref{sec.model} with a constant biomass conversion efficiency $\xi$), with $\xi \equiv 0.3$, mutation variances $\sigma_z=0.01$, $\sigma_\mu=0.001$ and parameters of Table~\ref{table.parametres}.
}
\end{figure}

\subsection{Threshold on $a$ for negative food webs}
\label{sec.simu.impact.a}

\begin{figure}
\begin{center}
 \includegraphics[width=14.6cm, trim = 3.4cm 4.8cm 3.5cm 3.8cm, clip=true]{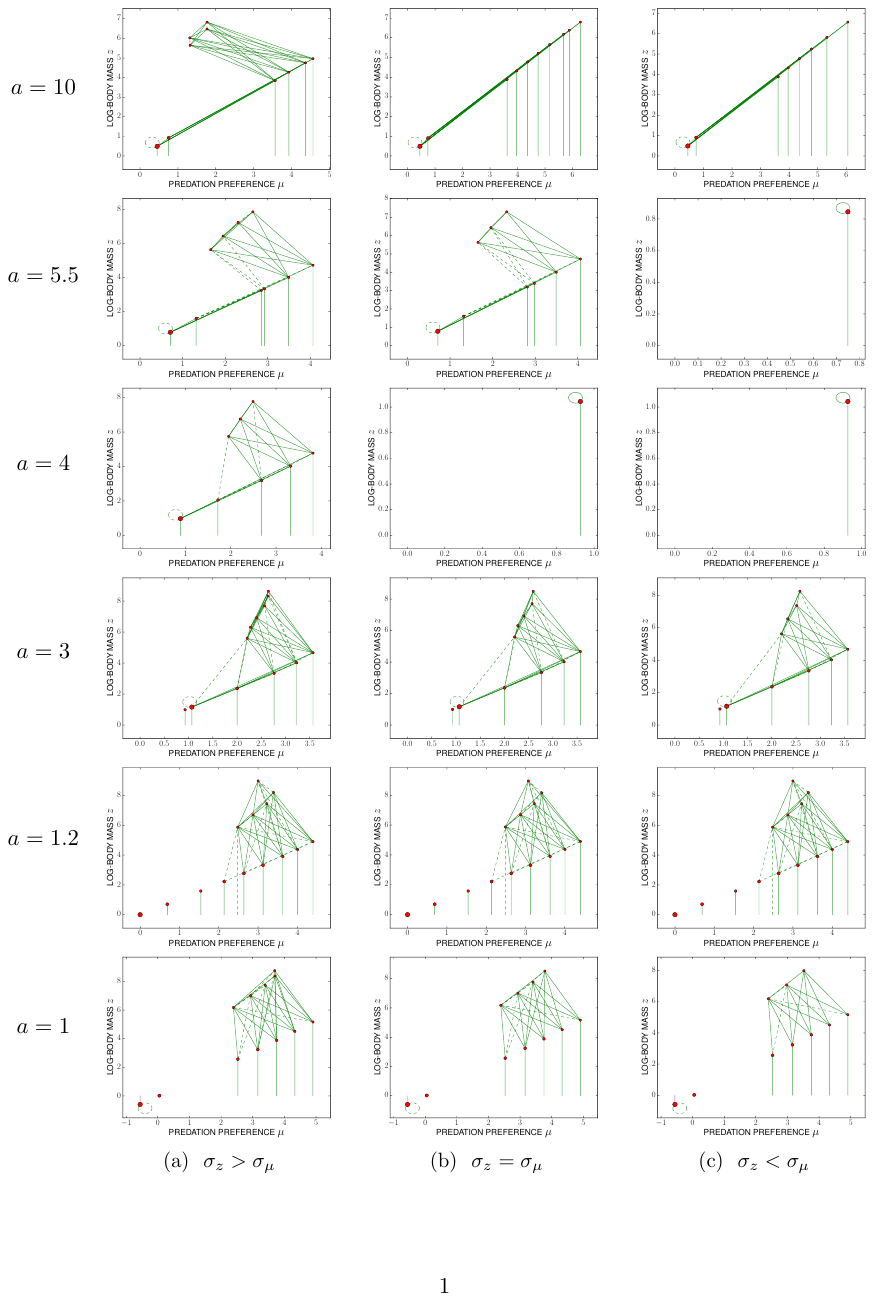}
 
\captionsetup[subfigure]{justification=centering}
\hspace{1.3cm}
\begin{subfigure}{4.4cm}
\caption{\label{fig.xi.lineaire.z_larger_mu} $\sigma_z>\sigma_\mu$}
\end{subfigure}
\begin{subfigure}{4.4cm}
\caption{\label{fig.xi.lineaire.z_eq_mu} $\sigma_z=\sigma_\mu$}
\end{subfigure}
\begin{subfigure}{4.4cm}
\caption{\label{fig.xi.lineaire.z_less_mu} $\sigma_z<\sigma_\mu$}
\end{subfigure}
\end{center}
\vspace{-0.5cm}
\caption{\label{fig.xi.lineaire}Food web at the stationary state (except for $a=5.5$ and
  $\sigma_z\geq \sigma_\mu$ and for $a=10$ and $\sigma_z> \sigma_\mu$ which produce oscillations, see
  Figure~\ref{fig.xi.lineaire.dynamics}) for different values of $a\geq 1$ and for mutation variances $\sigma_z=0.001<\sigma_\mu=0.01$ (a), $\sigma_z=\sigma_\mu=0.01$ (b) and
  $\sigma_z=0.01>\sigma_\mu=0.001$ (c). $b=0.15$ and $\xi_{\max}=0.75$. Other parameters are given in Table~\ref{table.parametres}.}
\end{figure}

\begin{figure}
\begin{center}
\includegraphics[width=14.8cm, trim = 3.3cm 6.3cm 3.5cm 5.2cm, clip=true]{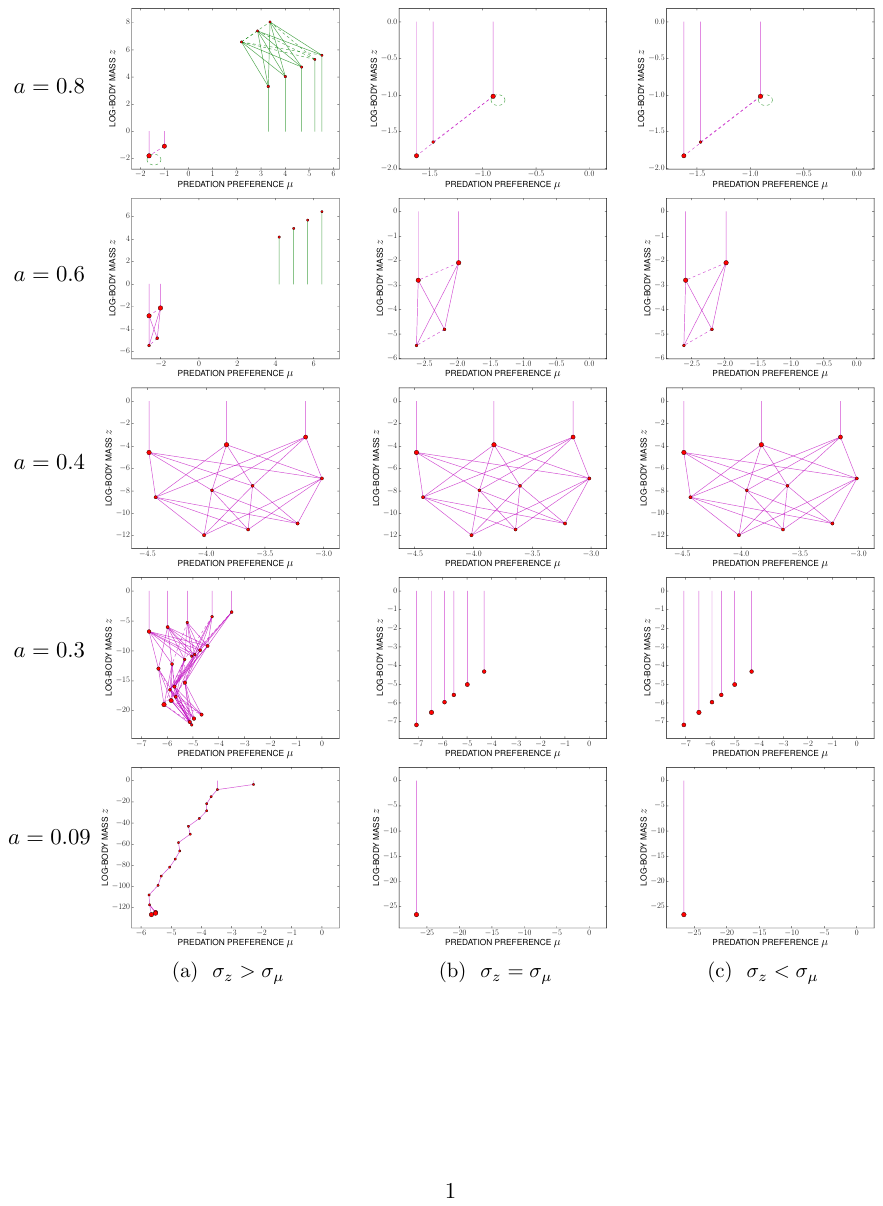}
 
\captionsetup[subfigure]{justification=centering}
\hspace{1.3cm}
\begin{subfigure}{4.4cm}
\caption{\label{fig.xi.lineaire.neg.z_larger_mu} $\sigma_z>\sigma_\mu$}
\end{subfigure}
\begin{subfigure}{4.4cm}
\caption{\label{fig.xi.lineaire.neg.z_eq_mu} $\sigma_z=\sigma_\mu$}
\end{subfigure}
\begin{subfigure}{4.4cm}
\caption{\label{fig.xi.lineaire.neg.z_less_mu} $\sigma_z<\sigma_\mu$}
\end{subfigure}
\end{center}
\caption{\label{fig.xi.lineaire.neg}Food web at the stationary state (except for $\sigma_z>\sigma_\mu$, $a=0.09$) for different values of $a<1$ and for mutation variances $\sigma_z=0.001<\sigma_\mu=0.01$ (a), $\sigma_z=\sigma_\mu=0.01$ (b) and  $\sigma_z=0.01>\sigma_\mu=0.001$ (c). $b=0.15$ and $\xi_{\max}=0.75$. Other parameters are given in Table~\ref{table.parametres}.}
\end{figure}

The arguments of Section~\ref{sec.a=1} suggest that species with negative phenotypes could emerge in the food web only if $\xi'(0)/\xi(0) = a\lessapprox 1$. This is confirmed
by the numerical study shown in Figures~\ref{fig.xi.lineaire} and~\ref{fig.xi.lineaire.neg}.


These figures show typical food web structures at the stationary state (except for
four particular simulations, see legend and Appendix~\ref{app.fw.dynamics}) for several values of $a$ (in rows) and several relative speeds of evolution
$\sigma_z$ and $\sigma_\mu$ of the two traits (in columns). Figures~\ref{fig.xi.lineaire} and~\ref{fig.xi.lineaire.neg} show that the
 variation of reproduction efficiency, governed by parameter $a=\xi'(0)/\xi(0)$, strongly affects the structure of the food webs (compare rows), while the relative speed of evolution
generally has  little effect, except for particular values of $a$ (compare columns).   

When the body size evolves faster than the predation log-distance ({\it i.e.} $\sigma_z>\sigma_\mu$),
realistic food webs ({\it i.e.} non-negative and non-trivial) emerge when the slope of $\xi$
is large ($a\geq 1.2$): starting with a single heterotrophic species, the community diversifies and the trophic network evolves until
a steady state is reached, where all species have positive traits and some species feed mainly on other species and
  not on the autotrophic resource (Figures~\ref{fig.xi.lineaire.z_larger_mu} and~\ref{fig.xi.lineaire.neg.z_larger_mu}). When
$1.2\leq a<3$ the food web contains a resource-like species which evolves to a size and a preferred predation distance closer to zero
as $a$ decreases. The resource-like species thus suffers strong cannibalism. When $a<1.2$, the
resource-like species evolves to a negative log-mass and starts feeding on prey (the resource) larger than itself. Meanwhile, the
richness of the food web decreases progressively until $a\approx 0.6$. Finally, when $a\to 0$, the conversion efficiency $\xi$
converges to the constant function $b$. We observe that the positive part of the food web goes extinct and the food web diversifies with only species smaller than the autotrophic
resource (as in Figure~\ref{fig.branstromSC.evol_mu}). As predicted
  in Section~\ref{sec.a=1}, the threshold at which the food web becomes partly negative is $a\approx 1$. With the
  parameters in Table~\ref{table.parametres}, approximation \eqref{eq.approx.partialf} becomes $\partial_1 f(0,0)\approx
  2.66[a-1]b(N^*_0+N^*_1)+0.025$. In all simulations shown in Figures~\ref{fig.xi.lineaire} and~\ref{fig.xi.lineaire.neg},
   $0.025$ is negligible with respect to $2.66 b(N^*_0+N^*_1)$ (for example, for $a=1.2$, the latter is
  larger than 10). This confirms that the sign of $\partial_1 f(0,0)$ is most often the same as $a-1$, so that a resource-like species can evolve to a negative log-mass if $a\lessapprox 1$. In our numerical study, the fact that this threshold seems to be closer to 1.2 than to 1 can be explained by the approximations we made
    to derive \eqref{eq.approx.partialf} (see Appendix~\ref{app.slope.xi}).

When the body size evolves at the same rate or more slowly than the preferred predation distance ({\it i.e.} $\sigma_z \leq
\sigma_\mu$), the evolution of food webs generally show similar patterns than previously described when $\sigma_z>\sigma_\mu$ except
for some intermediate values of $a$ where there is no diversification, leading to a trivial food web
(Figures~\ref{fig.xi.lineaire.z_eq_mu} and~\ref{fig.xi.lineaire.z_less_mu} with $a=4$ and
$a=5.5$).
A detailed study of the fitness landscape shows that, when $a=4$ and $a=5.5$, the first
  diversification event in the food web is due to evolutionary branching along slow directional evolution ({\it sensu} \cite{ito2014a}).
  When mutations are too fast on trait $\mu$, evolution becomes too fast to allow evolutionary branching, and the system
  reaches an evolutionary stable strategy corresponding to a trivial food web (see Appendix~\ref{sec.firstbranching} for
  details).

\subsection{Threshold on $b/\xi_\text{max}$ for trivial food webs}
\label{sec.simu.impact.b}

\begin{figure}
\begin{center}
 \includegraphics[width=15cm, trim = 3.4cm 7.5cm 3.5cm 7.2cm, clip=true]{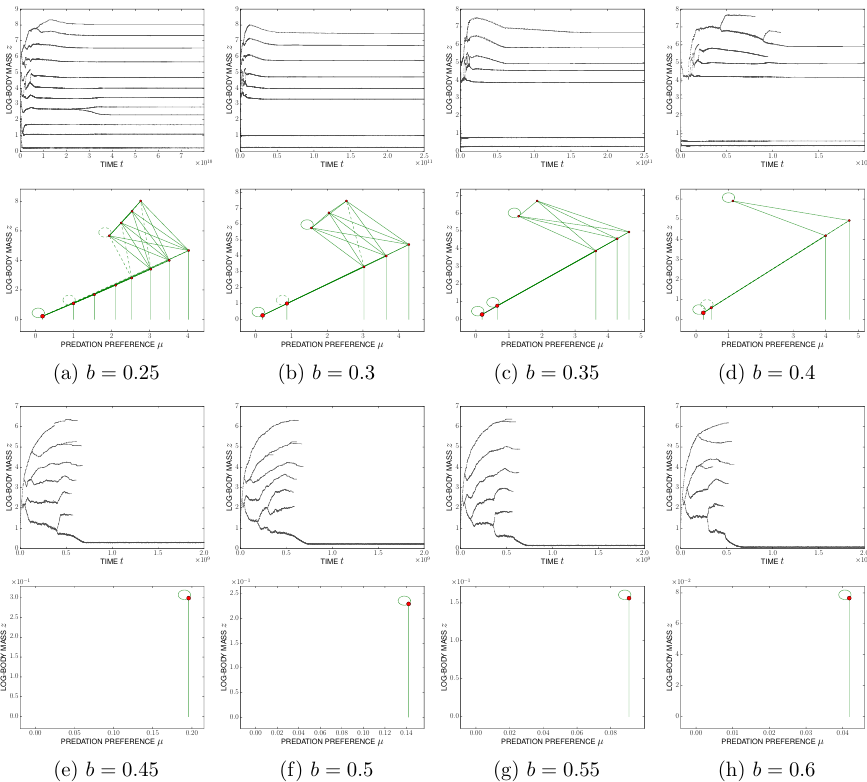}\\
\end{center}
\caption{\label{fig.b}Evolution of the log-mass $z$ (top) and food web at the stationary state (bottom) for $b$ varying from 0.25 to 0.6
with mutation variances $\sigma_z=0.01>\sigma_\mu=0.001$, $a=1.2$ and $\xi_{\max} = 0.75$. Other parameters are given in Table~\ref{table.parametres}.}
\end{figure}

Figure~\ref{fig.b} shows that the food webs structure
strongly depends on the parameter $b=\xi(0)$, corresponding to the biomass conversion efficiency  when a predator feeds on a prey with an identical log-body mass. For all values
of $b$, the initial food web dynamics shows progressive diversification as in Figure~\ref{fig.xi.lineaire.z_larger_mu}, but when the
smallest species becomes too close to 0, different behaviours are observed. When $b$ is small, the food web stabilises in a relevant
pattern. When $b$ increases, the richness of the food web progressively decreases: species with intermediate body masses go extinct,
until $b=0.4$ where only three species with large body masses remain. For $b\geq 0.45$, only the resource-like species survives and no
further diversification occurs. This can be understood from the argument
  of Section~\ref{sec.sensitivity.b}: in the situation where few species are present in the food web, among which a resource-like species,
  the resource-like species competitively excludes the others when the criterion~\eqref{extinction.criteria.xi.rls} is satisfied.
  With the chosen parameterization of $\xi$, $\xi(0,0)/\xi(z,0)\approx b/\xi_\text{max}$ for all sufficiently large log-body masses $z$ of non
  resource-like species. Using parameters in Table~\ref{table.parametres} and the observed trait values for the last surviving non
  resource-like species in Figure~\ref{fig.b}, we obtain an approximate threshold $b/\xi_\text{max} \gtrapprox 0.41$ for the
  emergence of trivial food webs, which is consistent with our simulations (see Appendix~\ref{app.b.xi} for computation details).

\section{Discussion}
In this paper, we analyse a unifying model of food webs adaptive
evolution that embraces most of the peculiarities of the
eco-evolutionary models which followed the seminal work by
\cite{loeuille2005a}. We introduce in particular a non-constant biomass
conversion efficiency as suggested by data and physical and behavioural
models. We show that relaxing the arbitrary constraints that are often 
assumed in the literature can generally yield unrealistic food webs
structures ({\it e.g.} relaxing the assumption of fixed preferred
predation distance in \cite{brannstrom2011}, or the constraints on
parameters range values in \cite{allhoff2015a,allhoff2016a}). We show
that assuming that the biomass conversion efficiency depends on prey and predator log-sizes restores the possibility of realistic food webs topologies
without the need of arbitrary constraints: body sizes of the species
remain larger than the basic resources size, and trophic levels emerge.
We also show that trophic networks evolution strongly depends on the
assumed form of the biomass conversion efficiency. \\

There are two manners to interpret our results. On the positive side, we
can conclude that our results bring a lot in identifying the
key mechanisms underlying the evolution of food webs. Indeed, we showed
that simply considering a non-constant biomass conversion
efficiency can solve most problems encountered with previous models. In
addition, the functional and behavioral ecology literatures suggest that there is indeed an
optimal prey size where predation is the most efficient
\citep{baras2014a, norin2017a, ho2019, pawar2019, portalier2018}. Our
model is, in a sense, closer to
data and observations, one might thus argue that such adaptive dynamics
models really tell us how food webs emerge from
eco-evolutionary processes. We can then go further into the analysis of
our results and discuss the importance of the biomass
conversion efficiency and its form. We numerically explored a single
family of functions for the biomass conversion efficiency
$\xi$ (linear w.r.t.\ the size difference between prey and predator and
truncated above and below fixed thresholds), but the fitness analysis
shows that we can extend
   the observed results to more general functions. We exhibited the effect
of two parameters on the expected trophic networks (Fig.\
\ref{fig.xi.lineaire}, \ref{fig.xi.lineaire.neg} and
\ref{fig.b}): the slope of biomass conversion efficiency at $0$ ({\it i.e.}\
$\partial_1\xi(0,0)/\xi(0,0)$) and the relative conversion efficiency $\xi(0,0)/\xi_{\text{max}}$, (Eq.\ \eqref{ex.xi1}, Fig.\ \ref{fig.ex.xi1}). Both parameters have threshold values at
which networks evolve such as either all species have a size smaller
than the resource, or the network is not stable and only one species
remains.
Assuming that the biomass conversion efficiency $\xi$ is constant ({\it
i.e.} $\partial_1\xi(0,0)/\xi(0,0)=0$) leads to resource-like species converting more energy into reproduction
as they feed on larger
prey.
Mutants decreasing size are thus necessarily favored and invade. Assuming a
non-constant biomass conversion efficiency $\xi$, depending on the
 log-sizes of prey and predator species, our
results show that: 1) $\xi(0,0)/\xi_{\text{max}}$ must be low enough to avoid the evolution to a
   network with a single resource-like species, and 2) $\partial_1\xi(0,0)/\xi(0,0)$ must be
large enough to avoid too large benefits for resource-like species to feed on
   larger prey and to avoid the evolution to a network with species
smaller than the resource.
In line with physical and behavioural models \citep{ho2019, pawar2019,
portalier2018}, our results suggest that mechanisms underlying how
energy is transferred from prey to predators, and how it is converted
into predators biomass, are key for the evolution and stability of food
webs.\\

However, one can interpret our results on a more negative side since they can suggest that such models are not robust enough
to gain informative insights about food webs evolution for three main reasons. First because we
show that relaxing arbitrary hypotheses can lead to the emergence of
qualitatively different and unrealistic networks topologies. Such models
are thus very sensitive to
assumptions and lack robustness. We showed that assuming a non-constant
biomass conversion efficiency can solve problems but one can argue that such
biomass efficiency (as well as many other parameters of the model) can
also evolve at the same time than body size and predation distance,
which could yield
to different food webs evolution. Second, because one would expect that
such models would be self-sufficient regarding
certain general properties of trophic networks. In particular, if size
effectively structures trophic networks with large species
feeding on smaller ones, and if cannibalism is rare, this should emerge
from the trait evolution and should not be due to arbitrary
constraints. The model analysed by \cite{brannstrom2011} does not assume
arbitrary constraints on the parameters range values, yet
its results are not robust to the hypothesis of fixed preferred predation
distance. Other models which avoid cannibalism by {\it a priori}
excluding it \citep{loeuille2005a, ritterskamp2016a}
obtain hierarchically structured trophic networks by assuming that
species can not feed on larger prey \citep{loeuille2005a}, assume
large mutations to observe network emergence \citep{allhoff2013a} or
arbitrarily constrain parameters range values in order to obtain
non-degenerated networks (Tab.\ \ref{table.models}). A general
robustness analysis of such models appears necessary before
generalising their results, such as the one performed by
\cite{brannstrom2011}. Finally,  we show that, even after
   introducing appropriate forms of the conversion efficiency and in
agreement with other authors, emerged food webs are very sensitive to
mutation sizes and the speed of evolution, the number of evolving traits
and the strength of interference competition, {\it i.e.}\ competition
which is not due to
resources and prey consumption. In particular, interference competition
seems unreasonably necessary for the evolution of food webs: it is not
clear whether there are good reasons why a non-trophic ecological
process would be so important in food webs evolution, especially in
models mostly based on competition for resources. Hence, our results
altogether with results by previous authors can cast doubts on
the explanatory and predictive power of such models. \\

One of the most important problems with the interpretation of such models
is that it is difficult to avoid circular reasoning. In most
previous papers, the authors claim that they are satisfied with the
model's results {\it because} they give realistic or relevant food webs
structures. Some authors even justify arbitrary assumptions, such as a
limited range for possible parameters values
\cite[\textit{e.g.}][]{allhoff2013a}, because otherwise results are not
satisfying.  We  use ourselves such a circular reasoning to justify
that assuming a non-constant biomass conversion efficiency $\xi$
actually solves many problems encountered by the models, which is not a
satisfying argument. It is
necessary to find ways to evaluate these models in a non-circular
manner. A possibility would be to identify quantitative independent
predictions that could be compared with food webs features. For
instance, some macroscopic properties are shared among observed networks
such as the distribution of the body log-mass ratio between prey and
predators. Considering non-constant biomass conversion efficiency yields
distribution which are close to the ones observed in empirical networks (see Fig. \ref{fig.size.distrib} in App. \ref{app.size.distrib})
which suggests that biomass conversion is indeed a key parameter
underlying food webs evolution. Another direction for testing the
validity of models would be to consider the limitation of the current
models which {\it a
   priori} assume allometric relationships between size and parameters,
generally as a power of $1/4$ \citep{peter1983a}. However, one
should observe that allometries measured in natural populations are {\it
evolved} values and not fixed parameters. In models,
including ours, allometric relationships are input parameters while they
should be output parameters. We suggest that a possible way
to evaluate such models would be to compare output allometric
relationships to observed data as already done by \cite{loeuille2006a}.

\section*{Speculation}

Our analysis and conclusions suggest that such adaptive models of food webs evolution might have reached their epistemological limits. If so, the main question remains: how and why do trophic networks emerge and evolve? Addressing this question might need reconsidering the category of models used by theoretical ecologists. What should then be the next epistemological step? One of the biggest limit of models based on Lotka-Volterra dynamical systems is that they are macroscopic, while we want to model interactions between individuals, i.e. ecological interactions at the microscopic level. A possible direction for future models dealing with food webs evolution might be to explicitly formalize how individuals interact. This could be done using agent-based simulations, spatialized or not, or simple probabilistic models. Following such a direction would in addition need to fill the gap between behavioural, functional, community and evolutionary ecologists. Indeed, functions such as ``predation rate'', ``reproduction rate'', ``functional responses'', ``allometries'', etc. would need to directly emerge from the interactions between individuals, and not to be phenomenological only, as they actually are in models, including ours.

\section*{Acknowledgements}
This work was supported by the Chaire ``Mod\'elisation Math\'ematique et Biodiversit\'e'' of VEOLIA Environment, \'Ecole Polytechnique, Mus\'eum National d'Histoire Naturelle and Fondation X. 
Simulations are run on the \textit{babycluster} of the \textit{Institut \'Elie Cartan de Lorraine}~: \url{http://babycluster.iecl.univ-lorraine.fr/}.

\appendix

\section{Detailed computation for the emergence of negative food webs}
\label{app.slope.xi}

Since we consider small mutations, the direction of
evolution of a given species $(z,\mu)$ is governed by the fitness gradient $\nabla_{(y,\eta)}f(y,\eta)|_{(y,\eta)=(z,\mu)}$
\citep{metz1996a, geritz1998a, dieckmann1996a, champagnat2011a, champagnat2001a}, where
\begin{align}
\partial_y f(y,\eta)
\nonumber	&=
		\sum_{i=0}^{n} \left[\partial_y\lambda(y,z_i) -\frac{y-z_i-\eta}{\sigma_\gamma^2}\,\lambda(y,z_i)\right]\,
		\gamma(y-z_i-\eta)\,N_i^*
\\
\label{df.y}
	&\quad
		-\sum_{i=1}^{n} \frac{z_i-y-\mu_i}{\sigma_\gamma^2}\,\gamma(z_i-y-\mu_i)\,N_i^*
		+\sum_{i=1}^{n} \alpha'(z_i-y)\,N_i^* - m'(y)
\end{align}
and
\begin{align}
\label{df.eta}
\partial_\eta f(y,\eta)
	&=
		\sum_{i=0}^{n} \frac{y-z_i-\eta}{\sigma_\gamma^2}\,
		\lambda(y,z_i)\,\gamma(y-z_i-\eta)\,N_i^*\,.
\end{align}
If $\xi$ is constant then $\partial_y \lambda(y,z_i)=-\lambda(y,z_i)$, it follows from these expressions that a specialist species ({\it i.e.}\ a species whose growth is mainly due to a
single prey size) has a tendency to evolve toward smaller traits. Indeed, for such a species $(z,\mu)\in (z_i,\mu_i)_{1\leq i \leq n}$, assuming that its major prey all have log-mass $\tilde z$ with total density $\tilde N^*$ (which is the sum of the resource density and the density
  of the resource-like species when $\tilde{z}=0$)
\[
\partial_\eta f(y,\eta)|_{(y,\eta)=(z,\mu)} \approx\frac{z-\tilde z-\mu}{\sigma_\gamma^2}\lambda(z,\tilde z)\gamma(z-\tilde z-\mu) \tilde N^*,
\]
which makes the trait $\mu$ evolve to $z-\tilde z$. Provided $\mu$ evolves fast enough,  we can assume that $\mu\approx z-\tilde z$, and then
\begin{align*}
\partial_1 f(z,\mu)=\partial_y f(y,\eta)|_{(y,\eta)=(z,\mu)}\approx & \partial_1\lambda(z,\tilde z)\,\gamma(0)\,\tilde N^*-\sum_{i=1}^n \frac{z_i-z-\mu_i}{\sigma_\gamma^2}\,\gamma(z_i-z-\mu_i)\,N_i^* \\
		& +\sum_{i=1}^{n} \alpha'(z_i-z)\,N_i^* - m'(z).
\end{align*}
If we assume in addition that all species $(z_i,\mu_i)$ predating $(z,\mu)$ ({\it i.e.}\ such that $\gamma(z_i-z-\mu_i)$ is not negligible)
are also such that $\mu_i\approx z_i-z$, and that other species than $(z,\mu)$ have different enough sizes so that the main part of 
competition acting on $(z,\mu)$ is exerted by species $(z,\mu)$ itself, we obtain (using that $\alpha'(0)=0$)
\[
\partial_1 f(z,\mu)\approx\partial_1\lambda(z,\tilde z)\,\gamma(0)\,\tilde N^* - m'(z).
\]
If $\tilde N^*$ is large enough, the last quantity has the same sign as $\partial_1\lambda(z,\tilde z)$, which is negative if $\xi$ is
constant. If $\xi$ is not constant and such as $\partial_1\lambda(z,\tilde z)$ is positive, this trend is reversed.

This argument applies in particular to cases with a resource-like species, {\it i.e.}\ a species with trait $(z,\mu)\approx(0,0)$. In the
simulations of Figures~\ref{fig.xi.lineaire}, \ref{fig.xi.lineaire.neg} and~\ref{fig.xi.lineaire.dynamics}, the loss of richness of
the food web and the appearance of unrealistic (negative) patterns seem closely related to the situations where a resource-like
species evolves to negative sizes. In these simulations, at the time where the resource-like species crosses $(0,0)$, the resource
and resource-like species can be considered as a single species, which is
relatively far from the rest of the food web, so that competition from other species is negligible and predation is exerted on this
species only from specialist species. Therefore, letting 1 be the index of the resource-like species
\begin{align*}
\partial_1 f(0,0)
	&\approx
		 \partial_1\lambda(0,0)\,\gamma(0)\,(N_0^*+N^*_1)  - m'(0).
\end{align*}

\section{Detailed computation for the emergence of trivial food webs}
\label{app.b.xi}
We assume that the food web is composed of the resource, a resource-like species for which we shall assume for simplicity
that trait is $(y,\eta)=(0,0)$ ({\it i.e.}\ resource consumption is optimal for this species), and a second species
with traits $z$ and $\mu$.

In this case, we shall use the competitive exclusion principle to decide whether the species $(z,\mu)$ is excluded by the resource like-species $(0,0)$. This will occur if the fitness $f(z,\mu)$ of the species $(z,\mu)$ is negative, where
\begin{equation}
  \label{eq:fitn-b-xi}
  f(z,\mu)=\lambda(z,0)\gamma(z-\mu)(N^*_0+N^*_1)-\alpha(-z)N^*_1-\gamma(-z)N^*_1-m(z),  
\end{equation}
and $N^*_0$ and $N^*_1$ are the equilibrium densities of resource and resource-like species respectively, when
the species $(z,\mu)$ is extinct. Hence for $\lambda(0,0)$ not too small (so that $N_1^*>0$)
\[
N^*_1=\frac{\lambda(0,0)\gamma(0)\frac{r_g}{k_0}-m(0)}{\frac{\lambda(0,0)\gamma(0)^2}{k_0}+(1-\lambda(0,0))\gamma(0)+\alpha(0)}\quad\text{and}
\quad N^*_1 + N^*_0=\frac{m(0)+(\gamma(0)+\alpha(0))N^*_1}{\lambda(0,0)\gamma(0)}.
\]
We then obtain
\begin{align*}
  f(z,\mu)
  &=  
  \frac{\lambda(z,0)\,\gamma(z-\mu)}{\lambda(0,0)\,\gamma(0)}\left[(\gamma(0)+\alpha(0))N^*_1+m_0\right]
  -(\alpha(-z)+\gamma(-z))\, N^*_1-m(z).
\end{align*}
Then the species $(z,\mu)$ cannot survive if
$$
	 \frac{\lambda(0,0)}{\lambda(z,0)} = \frac{\xi(0,0)}{\xi(z,0)\,e^{-z}}\geq 
		\frac{(\gamma(0)+\alpha(0))N^*_1+m_0}{(\alpha(-z)+\gamma(-z))\, N^*_1+m(z)}
		\frac{\gamma(z-\mu)}{\gamma(0)}.
$$

\medskip

This criteria explains the collapse of the food web observed in Figure~\ref{fig.b}, with $\xi$ defined by \eqref{ex.xi1}, for large values of $b$ when the trait of the smallest
species approaches $(0,0)$.
Observing that $\xi(0,0)=b$ and assuming that $a$ and $z$ are large enough so that $\xi(z,0)=\xi_{\max}$ (and indeed the
  observed values of $z$ are between 4 and 5 in simulations of
  Figure~\ref{fig.b}), we obtain, with the values of the parameters of Table~\ref{table.parametres},
\begin{align*}
N^*_1 
	&=  
	\frac{b\,2.66\frac{10}{0.01}-0.1}{\frac{b\,2.66^2}{0.01}+(1-b)\,2.66+0.8}
	\approx
	\frac{b\,2.66\frac{10}{0.01}}{\frac{b\,2.66^2}{0.01}}
	\approx 3.76,
\end{align*}
so that the species cannot survive if
$$
	\frac{b}{\xi_{\max}}\gtrapprox 
		\frac{4.93\,\gamma(z-\mu)e^{-z}}{3.76\,(\alpha(-z)+\gamma(-z))+0.1\,e^{-z/4}}
$$
Therefore, we obtain a threshold effect for large values of $b/\xi_{\max}$, and we indeed observe in the
simulations of Figure~\ref{fig.b} trait values  of the last
surviving non resource-like species  close
to $z=4$ and $\mu=2.8$, which gives an approximate threshold of $0.41$ for $b$,
above which we predict that the food web should collapse when the smallest species gets too close to trait $(0,0)$. This is
consistent with the simulations of Figure~\ref{fig.b}.
Biologically, when $b$ increases, the resource-like species is more adapted for the resources consumption and for cannibalism.
Then, for large $b$, the resource-like species is too competitive for other species to survive.

\section{Numerical methods and additional simulations}

\subsection{Food web simulation method}
\label{app.simu.method}
The details of the computational methods slightly vary among references. Here we applied the following scheme.

Between mutation events, we compute the solution of Eq.~\eqref{general.eq}-\eqref{eq.ressource} using the \texttt{odeint} python solver. Mutations are assumed to occur each $t_m=5.10^4$ time units. This gives similar, yet faster, results to alternative schemes where mutations occur with a small probability at each time steps. 
The species $(z,\mu)$ producing a mutant is drawn proportionally to its individual density.
The mutant $(z',\mu')$ is drawn such that $z'$ and $\mu'$ are independent and Gaussian with means $z$ and $\mu$ and variances $\sigma_z^2$ and $\sigma_\mu^2$ respectively. 
The mutant initial density is a small value $\varepsilon$ and
species are assumed to go extinct if their densities go below the same threshold $\varepsilon$.
All  simulations  were performed with $\varepsilon=0.0001$. The density of the initial species and the initial resource concentration are
the equilibrium of the system~\eqref{general.eq}-\eqref{eq.ressource} with $n=1$ given by
\begin{align}
\label{eq.equilibre.1pop}
	N_1^* = \frac{\lambda_{10}\,\gamma_{10}\,r_g-k_0\,m_1}
		{\lambda_{10}\gamma_{10}^2+k_0\,((1-\lambda_{11})\,\gamma_{11}+\alpha_{11})}\,,
		\qquad
	N_0^* = \frac{r_g-\gamma_{10}\,N_1^*}{k_0}
\end{align}
assuming that the fitness of species $1$ satisfies $\lambda_{10}\,\gamma_{10}\,r_g-k_0\,m_1>0$,
where $m_1$, $\lambda_{10}$, $\lambda_{11}$, $\gamma_{10}$ and $\gamma_{11}$ are the death rate, production efficiencies and
predation rates (here reduced to resource consumption and cannibalism) associated to the initial species.

\bigskip

The resulting food webs are represented with  an edge  drawn between predator $i$ and
prey $j$ (or a loop if $j=i$) if predation of $j$ by
$i$ (or cannibalism if $j=i$) is responsible for more than 10\% (5\% for dashed edges) of the reproduction of species $i$, {\it i.e.}
\begin{align}
\label{eq.reprod.larger0.1}
\frac{\lambda_{ij} \, \gamma_{ij}\,N_j}
{\sum_{k = 0}^{n} \lambda_{ik} \, \gamma_{ik}\,N_k}
	>0.1	\,.
\end{align}

Note that links between species are drawn based on reproduction rates $\lambda_{ij} \gamma_{ij}\,N_j$  rather than on predation rates only $\gamma_{ij}\,N_j$ because, due to the dependency of $\lambda_{ij}$ on the prey and the predator sizes,
even if a prey is little consumed compared to other prey, it may contribute to a large part of reproduction of its predator.
This is particularly true for large prey, for example with cannibalism.

\subsection{Regularisation of $\xi$}
\label{app.reg.xi}
To avoid problems of irregularity of fitness functions, we use in simulations a regularisation of the function $\xi$ defined by \eqref{ex.xi1}. The function $\xi$ can be obtained as an affine transformation of the function $\xi_0$ defined by
\begin{align*}
\xi_0(x) & =
	\begin{cases} 
	0 & \text{if $x \leq -1$}\\
	1+x & \text{if $-1\leq x \leq 1$}\\
	2 & \text{if $x \geq 1$}.
	\end{cases}
\end{align*}
The regularisation of $\xi$ is obtained with the same affine transformation applied to the following regularisation of $\xi_0$
\begin{align}
\label{ex.xi0.reg}
\tilde \xi_0(x) & =
	\begin{cases} 
	0 & \text{if $x \leq -1-\varepsilon$}\\
	\frac{(x+1+\varepsilon)^3}{6\,\varepsilon^2} & \text{if $-1-\varepsilon \leq x \leq -1$}\\
	-\frac{(x+1-\varepsilon)^3}{6\,\varepsilon^2}+x+1 & \text{if $-1\leq x \leq -1+\varepsilon$}\\
	(x+1) & \text{if $-1+\varepsilon \leq x \leq 1-\varepsilon$}\\
	-\frac{(x-1+\varepsilon)^3}{6\,\varepsilon^2} +x+1 & \text{if $1-\varepsilon \leq x \leq 1$}\\
	\frac{(x-1-\varepsilon)^3}{6\,\varepsilon^2} +2 & \text{if $1 \leq x \leq 1+\varepsilon$}\\
	2 & \text{if $x \geq 1+\varepsilon$}
	\end{cases}
\end{align}
with $\varepsilon=0.7$.

\subsection{Food web dynamics}
\label{app.fw.dynamics}

\begin{figure}
\begin{center}
 \includegraphics[width=15.3cm, trim = 3.5cm 6cm 3cm 5cm, clip=true]{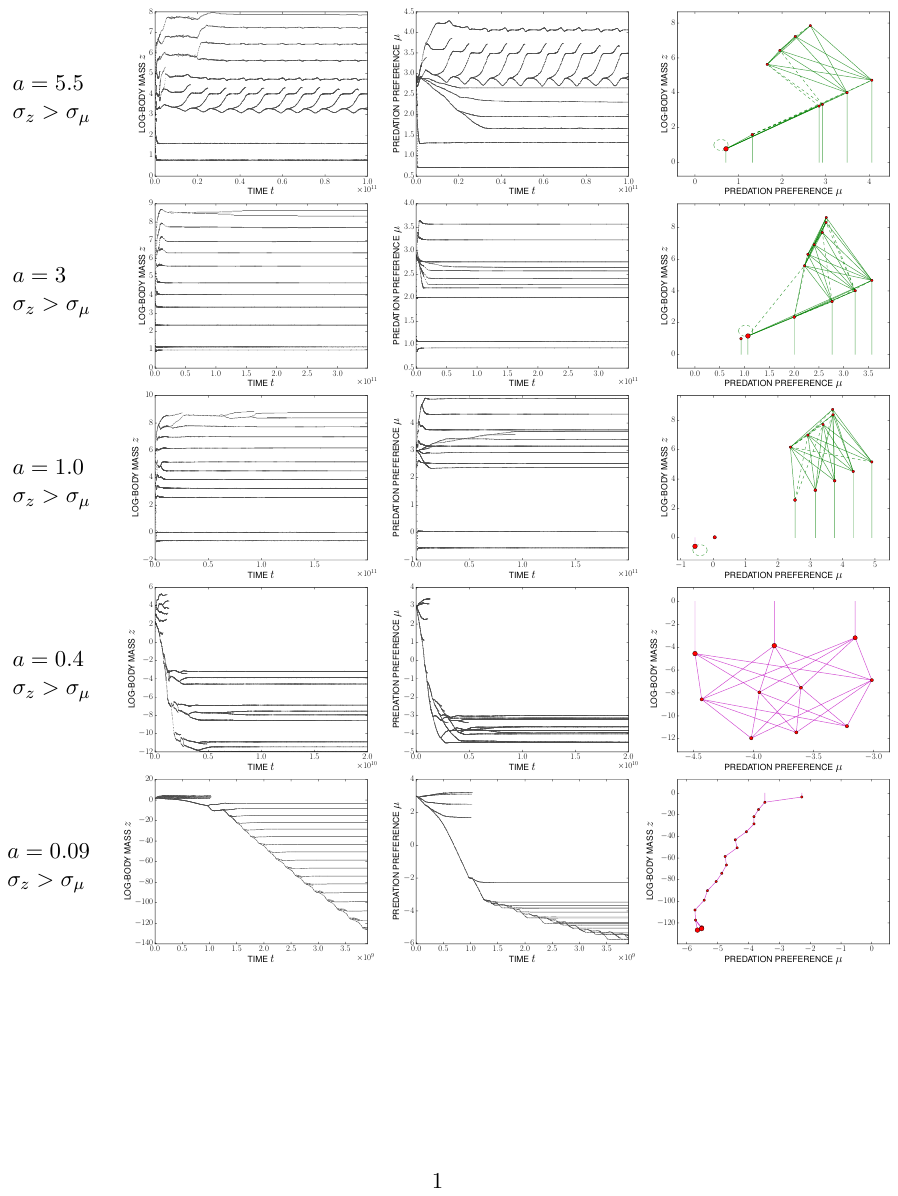}
\end{center}
\caption{\label{fig.xi.lineaire.dynamics}Evolution of the log-mass $z$ (left), evolution of the predation preference $\mu$ (middle)
  and the final food web (right), for $\xi$ defined by \eqref{ex.xi1} with $\xi_{\max}=0.75$, $b=0.15$, $\sigma_z=0.01>\sigma_\mu=0.001$ and $a$ varying between 0.09 and 5.5.}
\end{figure}

The dynamics of some simulations of Figures~\ref{fig.xi.lineaire} and \ref{fig.xi.lineaire.neg} are shown in Figure~\ref{fig.xi.lineaire.dynamics}. They confirm that the food webs
  shown in Figures~\ref{fig.xi.lineaire} and~\ref{fig.xi.lineaire.neg} are stationary, except for $a=5.5$ and
    $\sigma_z> \sigma_\mu$, where periodic dynamics occur in the evolution of both traits $z$ and $\mu$ (similar behaviour is
  observed for $a=10$ and $\sigma_z>\sigma_\mu$ and for $a=5.5$ and $\sigma_z=\sigma_\mu$). 
  The emergence of evolutionary cycles as observed for $a=5.5$ and $\sigma_z> \sigma_\mu$ was studied in \cite{ritterskamp2016b}. 
 In the simulation with  $a=0.09$ and $\sigma_z>\sigma_\mu$, the food web does
    not reach a stationary
    state either and seems to evolve endlessly to smaller negative log-masses as in Figure~\ref{fig.branstromSC.evol_mu}.   We also see that, for values of $a$ much smaller than $1$, the food web first evolves
  to a realistic shape, similar to those observed for larger values of $a$, until the smallest body mass becomes too negative. After
  this, the positive part of the food web collapses and only the negative part remains.

\subsection{Effect of the variance of the interference competition kernel}
\label{app.competition}

\begin{figure}
\begin{center}
 \includegraphics[width=15cm, trim = 3.5cm 12.3cm 3.5cm 12cm, clip=true]{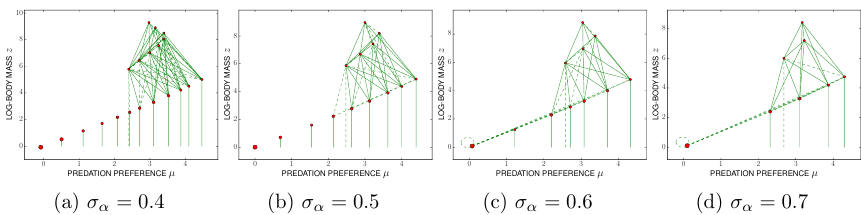}
\end{center}
\caption{\label{fig.sigma_alpha}Food web at the stationary state for different values of the competition kernel width (a) $\sigma_\alpha=0.4$, (b) $\sigma_\alpha=0.5$, (c) $\sigma_\alpha=0.6$, (d) $\sigma_\alpha=0.7$
with mutation variances $\sigma_z=0.01>\sigma_\mu=0.001$, $a=1.2$, $b=0.15$ and $\xi_{\max}=0.75$. Other parameters are given in Table~\ref{table.parametres}.}
\end{figure}

As observed previously \citep[\textit{e.g.}][]{brannstrom2011} the variance $\sigma_\alpha$ of the
competition kernel has a strong influence on the richness of the food web. Indeed, Figure~\ref{fig.sigma_alpha} shows that the smaller $\sigma_\alpha$, the richer the stationary food webs.


\subsection{Predicted distribution of prey {\it vs.} predator body size}

\begin{figure}
  \begin{center}
     \includegraphics[width=8cm]{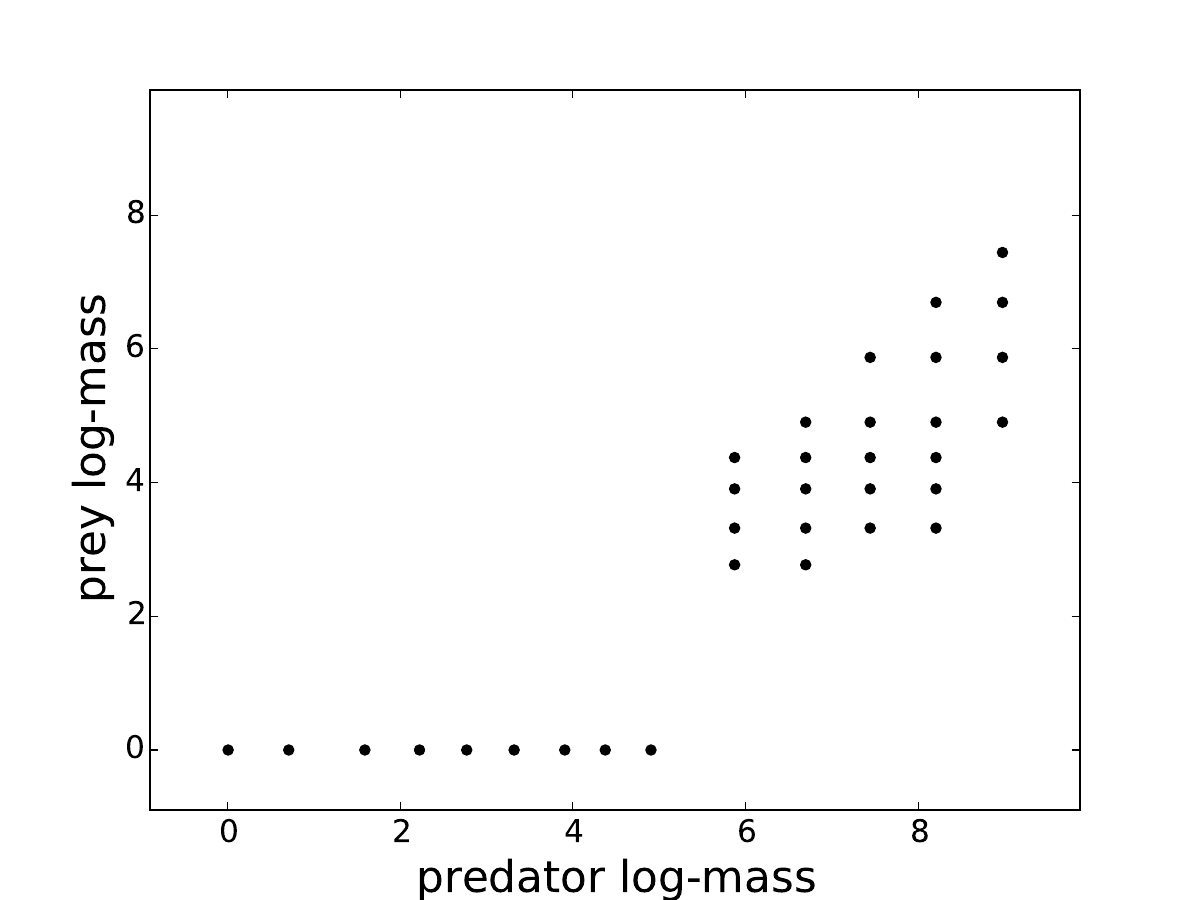}
\end{center}
\caption{\label{fig.size.distrib} Distribution of the predator {\it vs.} prey size in a stationary food web. A dot depicts a predation interaction between two species where the biomass flux between the prey and predator is significant. This distribution corresponds to the food web presented in Fig.\ref{fig.xi.lineaire} where $a=1.2$ and $\sigma_z>\sigma_\mu$.}
\end{figure}

Figure~\ref{fig.size.distrib} describes the distribution of the predator and prey sizes in the stationary food web presented in Figure~\ref{fig.xi.lineaire} with $a=1.2$ and $\sigma_z>\sigma_\mu$.
Even though the number of species in the food web emerging from our model is lower than in observed food webs, the distribution are very similar (compare for instance Figure 5 in \cite{ho2019}).

\label{app.size.distrib}
    
\section{A review of  food web evolution models following \cite{loeuille2005a}}
\label{sec.review}

Eco-evolutionary models of food web evolution derived from \cite{loeuille2005a} are inspired by the adaptive dynamics framework.
 Clonal species are supposed to evolve because of successive invasions of mutations introduced into a population at ecological equilibrium.
Ecological dynamics are given by a set of deterministic Lotka-Volterra equations, including resource consumption, predation,
competition, birth and death. A resource is also considered, with its own dynamics on while species may feeds and with input independent of the rest of the system. The main idea behind these models is that food web structure is driven by traits of the species, especially individuals' size or mass. Other traits are also considered, such as the preferred relative size of prey and niche width of predators.
In general, a higher dimensional trait space is known to promote branching \citep{ispolatov2015a,doebeli2017a}.
In all models, allometric scaling relationships are considered between one or several parameters, for instance individual death or reproduction rates, and the size or the body mass of individuals, following well-known empirical observations \citep{peter1983a}.

In all models, at least size (or body mass) is affected by mutations corresponding to the introduction of
new species (or morphs) in the food web and evolves, assuming separation of timescales between ecological
and evolutionary processes: new (mutant) species are introduced one after another on a long timescale, allowing ecological equilibrium to be reached between mutations. These models' assumptions and objectives are thus different than the ones studying communities and ecosystems stability at short ecological timescale \citep[\textit{e.g.}][]{brose2006a, otto2007a, allesina2012, miele2019}.
Mutant traits are drawn from a random distribution. A mutant species can be favored or not depending on the species already present in
the environment. If it is favored, the mutant species invades and either replaces resident species or coexist with at least one other species.  The analysis of the model is generally conducted using numerical simulations following a Piecewise-Deterministic Markov Process: population dynamics are deterministic and random mutations are introduced at random or fixed times.

All models assume that a single species is initially present in the environment, with the resources. Two phenomena can produce
species diversification and the emergence of a food web. When only small mutations are considered, the food web gets enriched
  by a new species through evolutionary branching ({\it sensu} Adaptive Dynamics). When large mutations are
possible, a new species may by chance appear in trait regions allowing invasion and coexistence with extant species, even though the parent species is locally stable (seeming rather like an immigration phenomenon). In all cases, models show that for a
given set of parameters, the number of species can increase. Trophic interactions between species evolve since species i) can feed on
other species, ii) have a preferred size of prey, iii) are more or less specialised on a prey size, iv) at least individuals'
size evolves. The results are hence generally presented as an interaction network, with the species as nodes, and trophic interactions between species as edges.  Finally, all models claim that for a set of reasonable parameter values, food webs can emerge and evolve, and their structure can be close to trophic networks encountered in nature.

In the following, we review adaptive evolution models of food webs derived from the seminal work by \cite{loeuille2005a}. Despite that these models share common foundations, some of their assumptions can substantially vary. Table~\ref{table.models} compiles the models we review here and gives details about how they differ and why it matters. 

{\bf Which traits evolve?} The number of evolving traits can differ, giving rise to a large variability of results. When only size
evolves \citep{loeuille2005a,brannstrom2011}, food webs are structured mostly linearly, like a linear food chain, and trophic levels are clearly defined. Similar results are obtained if size evolves
with other traits such that relative prey size preference or niche width \citep{ingram2009, allhoff2013a, allhoff2015a, allhoff2016a, bolchoun2017a}. However, in some cases food webs can show unrealistic
structures, {\it e.g.} with many species in a single trophic level, all feeding on the resource \citep{allhoff2013a}. In the work of \cite{ritterskamp2016a}, in addition to size, an abstract trait also evolves, which facilitates the emergence of food webs with different trophic levels and many different species. This shows that food web evolution largely depends on {\it a priori} assumptions about the evolving traits.

{\bf Mass/size or log mass/log size?} Models either assume that food webs are structured with individuals size/mass or log size/log mass.
The choice is important since, on the one hand, if the food web is structured linearly with
absolute size or mass, it implies that trophic interactions and predation rates depend on the absolute size difference between
predators and prey \citep{loeuille2005a, allhoff2013a}. If the predation distance (\textit{i.e.} the optimal predator-prey body mass difference) is not subject to mutation and is then the
  same for all species, relatively to their masses, small species feed on relatively much smaller species whereas large species feed on very
  similar species. Furthermore, for a given and fixed prey size preference, a small species cannot feed on any other species
  because no species could be small enough, hence giving artificial constraints to the minimal possible size of species. In addition,
  models considering a structure with the absolute size/mass also assume that the resource has a size/mass equal to 0, which prevents
  species to be smaller than the resource, a constraint that can be considered artificial. On the other hand, assuming log size/log
  mass \citep[as in][]{brannstrom2011, allhoff2015a, ritterskamp2016a, allhoff2016a} allows to describe trophic interactions and
  predation rates that depend on the relative
size between prey and predators through a predation log-distance (\textit{i.e.} the optimal predator-prey body mass ratio) rather than a predation distance. 
Models on the log-scale seem to be more realistic
  for food webs with fixed predation (log-)distance containing a large extent of (log) masses/(log) sizes. 
Moreover, models on the log-scale assume that the resource has a log size/log mass equals to 0 which allows species to
become smaller than the resource if log masses/log sizes are allowed to reach negative values. Overall, it appears that food web models would rather be structured following log size/log mass in order to introduce the least artificial constraints.

{\bf Boundaries.} Models make different implicit assumptions regarding the boundaries of possible values of evolving traits. As said
previously, assuming absolute rather than relative body size/mass inherently generates artificial boundaries. When other traits
evolve with body size/mass, {\it e.g.} with prey size preference or predation niche width, boundaries are also assumed. For example,
the predation log-distance and the niche width are constrained to a fixed interval in
\cite{allhoff2015a,allhoff2016a, bolchoun2017a} (see Table~\ref{table.models}). The reason why such boundaries are assumed is not
clear. It appears in some cases that if no boundary constraints are assumed, the models generally give aberrant or trivial resulting
trophic networks ({\it e.g.} \citealp{allhoff2013a} where final food webs contain only one hyper-specialised species or a single
trophic level; see also \citealp{allhoff2016a}). This problem has been solved by \cite{ingram2009} assuming that evolution of
  niche width is constrained by a trade-off on the predation rate. However, the chosen form of trade-off is hard to justify and the
  existence of a direct trade-off on niche width is not supported by biological observations, contrary to the trade-off on biomass
  conversion efficiency that we use.
If giving constrained boundaries to evolving traits is necessary to avoid unrealistic results, the validity of such a theoretical
framework can be questioned. On the other hand, one can expect that if such models capture the fundamental
mechanisms underlying the evolution of food webs, then traits should evolve to realistic values on their own, without being
artificially constrained.

{\bf Cannibalism, or not?} Cannibalism is excluded in some models, either because species are assumed to feed only on strictly
smaller species \citep{loeuille2005a,allhoff2013a}, or because an abstract trait is supposed as structuring the food web and
  species with identical body masses / sizes can feed on species which do not share the same abstract trait \citep{ritterskamp2016a}. {\it A priori} excluding cannibalism has several caveats. First, cannibalism is widespread in nature
  \citep{fox1975a}, and one can expect that models of food web evolution reflect all features of observed trophic networks,
  including cannibalism. Second, excluding cannibalism artificially constraints the fate of mutants. 
      For example, in the works of \cite{loeuille2005a} and \cite{allhoff2013a}, individuals with a very small size difference   can feed on
    and/or be eaten by each other while individuals with exactly the same trait can not.
    In \cite{loeuille2005a} this distinction has negligible effects since predation distance is fixed to a large
      value while  in \cite{allhoff2013a} this distance may evolve to small, positive values. Why biomass
  flows are possible between very similar (but different) species and not for identical
    species is not clear. 
	Furthermore, this introduces an additional selective pressure on neighbouring mutants: mutations close to their parent, which should be beneficial in cases with cannibalism (for instance by reducing direct competition), will usually be deleterious without cannibalism.
	Third, giving a justification for models excluding cannibalism based on
    individual-based models in the way of \cite{costa2016a} \citep[see
    also][]{champagnat2011a,takahashi2013a,champagnat2014a,campillo2015b} would require discontinuous predation kernels (with a
    different value when predating preys with exactly the same traits). This would introduce
    artificial singularities in fitness landscapes.

{\bf Ordered predation.} Some models \citep{loeuille2005a, allhoff2013a} {\it a priori} assume an order for predation: species can only feed on smaller species. It necessarily excludes cannibalism. Ordered predation also arbitrarily implies that in the case of species with very similar size, only the larger can feed on the smaller. Consequently, in the case of small mutations, it prevents trophic interaction changes: a prey can not become a predator. On the other hand, in the case of large mutations, a prey mutant can instantly become a predator if its size is larger than the size of the predator.
This also introduces an asymmetry in the fate of mutants: larger mutants are predators of their parent species whereas smaller mutants
are prey of their parent species, even though both species  may be very similar.
Overall, assuming an ordered predation kernel imposes a structure to the food web and strongly constrains the evolution of trophic interactions.

{\bf Mutation kernels.} Mutation kernels are assumed either Uniform or Gaussian, generally centred around the parents value, but not
necessarily. Mutations can have effect proportional to the parental value \citep{loeuille2005a,allhoff2013a} or independent of
parental value, sometimes mixing both assumptions if several traits evolve \citep{allhoff2015a,allhoff2016a}. Since models can make very different assumptions regarding their mutation kernels, it is
difficult to compare them. For instance, assuming mutations not centred on the parental value \citep{allhoff2015a} mimics migration
from a regional pool of species rather than mutations. 

{\bf Mutation size.} All models, except the one of \cite{brannstrom2011}, assume that mutation effects are large. Some authors even
show that if mutations are too small then diversification cannot occur \citep{allhoff2015a}. If mutations should have large effects, then evolutionary branching is certainly not an
important phenomenon in the evolution of trophic networks, since it requires small mutations in the adaptive dynamics
  framework.
  However, the model by \cite{brannstrom2011} showed that small mutations can be sufficient to make a relevant trophic
  network emerge. 
Overall, it is neither clear why small mutations are sufficient under some conditions and not under others, nor whether the assumptions about the mutation kernel can strongly affect the evolution of food webs.

{{\bf Interference competition.} In all models we reviewed a direct competition between species is added. Direct competition does not depend on trophic interactions but is supposed to be due to other mechanisms such as competitive interference, for instance for space. 
 Inter and intraspecific interference competitions are generally identified as
  important driving forces behind food web evolution because of two main reasons \citep[see \textit{e.g.}][]{allhoff2015a}. First, strong intraspecific
  interference competition forces populations to stay small, allowing more species to co-exist on the same amount of resource. This leads to
  higher diversity and larger networks.  Second, strong  interspecific interference competition induces similar species to evolve to different trait values hence allowing the emergence of different feeding strategies. Strong competition also affects the stability of food webs, as observed by \cite{ritterskamp2016b}: it prevents the
  emergence of complex  evolutionary dynamics, such as oscillating or even chaotic dynamics.
  However, the
  parameterization of this important mechanism is tricky. Usually, competition follows either a
  Gaussian function of the difference of traits, or an indicator function of an interval (\emph{i.e.} a constant function truncated out of this interval), most of the time without real
  justification. It may also be included in various ways in the functional response (additive factor, Beddington-DeAngelis, etc.). However, it is well-known that the shape of a competition kernel strongly influences the evolutionary patterns one
  may observe, specifically regarding the structuring of the population into well-separated sub-populations, interpreted here as
  distinct species \citep[cf. \textit{e.g.}][]{leimar2008a,  genieys2009a}. Therefore, the choice of a competition
  model is of great importance to shape food webs. Some attempts were done in \cite{allhoff2015a,ritterskamp2016a} to justify the
  chosen form of interference competition kernels, but this issue remains largely open. }

{\bf Variations of other features.} Models can also differ for other features, more or less importantly. \cite{loeuille2006a} studied
the emerging allometries between species densities and body masses in the model by \cite{loeuille2005a}. They concluded that the
exponent of allometry is strongly influenced by predation parameters. \cite{ritterskamp2016c} considered two resources;
\cite{guill2008a} assumed a variation of the form of the Beddington-DeAngelis functional response; \cite{allhoff2015a} and \cite{ingram2009} extended \cite{loeuille2005a}'s model to evolving feeding niche width and include a Beddington-DeAngelis functional
response. We also did not review models assuming that the food web is structured on another trait than size or body mass:
\cite{drossel2001a,drossel2004a,rossberg2008a,takahashi2013a} assumed an abstract discrete binary trait governing morphology,
behaviour or predation interactions {\it via} foraging and vulnerability. We invite the reader to read the above-cited
  references for further details. Even though these variations in assumptions seem minor at first glance, it is not clear whether or not they can significantly affect food webs evolution. However, for simplicity, we did not include these minor variations into our analysis. 

{\bf Biomass conversion efficiency.} Surprisingly, a single feature is common among all models: the \textit{biomass conversion efficiency} (or conversion factor), {\it i.e.} the fraction of ingested biomass devoted to the production of biomass of newborns, is supposed constant. Assuming that the conversion efficiency is independent of the mass / body size of predators and  prey is undoubtedly important because it implies that the mass converted by individuals when consuming prey is increasingly large when feeding on larger prey, with no limit. In other words, there is no cost for predation: feeding on larger prey is not more costly than on smaller prey. This is in contradiction with empirical data which show that there is a trade-off for predators between eating small and large prey \citep{baras2014a, norin2017a}: there is an optimal prey size for which the conversion efficiency is the highest. This can be due to a trade-off between the low biomass given by small prey but lower costs to forage, handle and digest than for larger prey. It can also be due to the fact that predators can feed only on a part of a large prey: the biomass converted from a large prey attains a maximum. This is also in contradiction with another common assumption of these models: if there exists a preferred size for prey, and if this preference can evolve, it should correspond to the best compromise between eating small or large prey. How the efficiency of biomass conversion affects the evolution of food webs has been ignored by extant models. We  show in the following that it is certainly a major mechanism underlying the evolution of trophic networks.

\bigskip

Overall, synthesising the common and different assumptions of the models of food webs evolution derived from \cite{loeuille2005a}
shows that the important mechanisms underlying the diversification and evolution of trophic networks are not clear. It is not clear
whether there is a single major mechanism that could give rise to relevant structured networks, or whether combinations of
different assumptions can give rise to similar results. This statement can cast doubt on the significance of such models for the
study of food web evolution because model construction, analysis and exploration can be biased by \emph{a priori} expectations of
network structure and general features. This is not necessarily because some combinations of assumptions give rise to relevant
network structure that the underlying mechanisms are correct. At best, such models can give directions to what should be empirically
verified: for instance, is it true that biomass conversion efficiency is constant? At worst, relaxing some of the assumptions made by
the extant models, especially artificial constraints such as excluding cannibalism or imposing boundaries to evolving traits, could
totally change the models' results and give aberrant, trivial or unrealistic network structure. This is what we will explore in the
following. We first propose a general model, unifying a large range of extant models. Second, we use our model to explore the results given by extant models when relaxing some of their assumptions. Finally, we use our model to explore the importance of the conversion efficiency.

\section{Emergence of unrealistic food webs under relaxed constraints}
\label{sec.relaxConstraints}
\begin{figure}
\begin{center}
 \includegraphics[width=7cm]{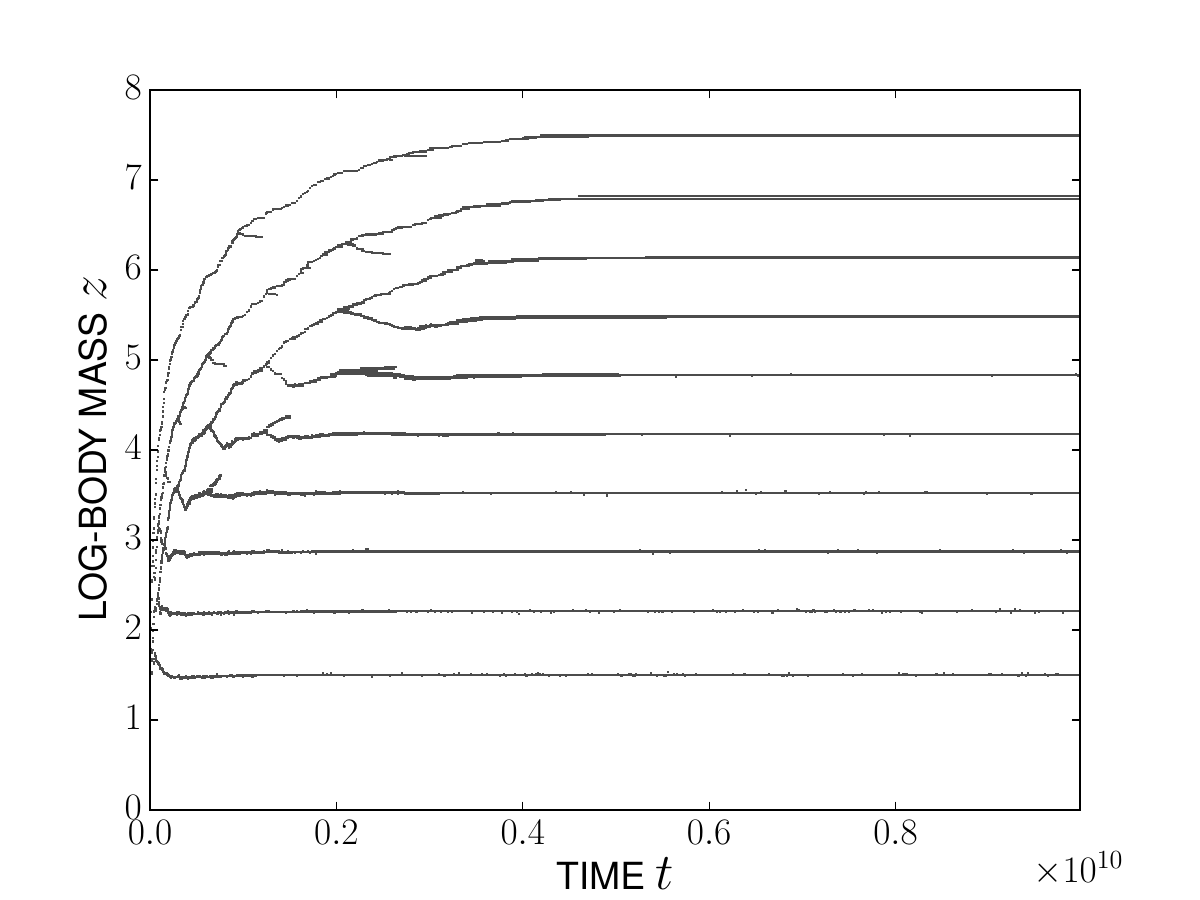}
\end{center}
\caption{\label{fig.branstrom}Evolution of the log-mass in the model by \cite{brannstrom2011} with a single initial species with log-body mass $z=1.2$, variance of the mutation distribution $\sigma_z=0.01$, predation preference $\mu=3$, biomass conversion efficiency $\xi \equiv \lambda_0 = 0.3$ (other parameters are given in Table~\ref{table.parametres}).
}
\end{figure}

In line with \cite{brannstrom2011}, with log-mass as the unique evolving trait, we obtain diversification by branching (see
Figure~\ref{fig.branstrom}) for identical parameters (given in Table~\ref{table.parametres}), except for the range of competition $\sigma_\alpha$ which is a bit smaller in order to favour branching events (see Figure~\ref{fig.sigma_alpha}).

Figure~\ref{fig.branstrom.evol_mu} shows three simulations where $z$ and $\mu$ can both evolve, with much smaller mutations on $\mu$ than $z$. 
In Figure~\ref{fig.branstrom.evol_mu.BSC}, the food web initially evolves as expected: several branching occur and the food web gets structured. However, the smallest species progressively evolves to smaller body size and predation preference until they both become negative.
This means that this species feeds on a larger prey: the resource. After this, the richness of the positive part of the food web ({\it i.e.} the network composed of species with positive body masses and positive predation preferences) decreases and the negative part of the food web progressively diversifies, producing a linear food web with more and more negative traits. 
The stationary state of the food web is not reached at the end of the simulation, as the food web seems to evolve similarly endlessly.

We ran simulations imposing positivity constraints on $z$ and $\mu$ (by truncating mutation distributions below 0) and obtained Figure~\ref{fig.branstrom.evol_mu.BAC}. 
The behaviour of the food web is similar to the one of Figure~\ref{fig.branstrom.evol_mu.BSC} until the species with the smallest body mass reaches zero.
After this time, this body mass remains close to zero and the predation preference goes to zero. This produces a progressive loss of species ending finally with a single resource-like species ({\it i.e.} a  species whose the size is close to the size of the resource species), with body mass and predation preference close to zero.
Again the behaviour is unrealistic and occurs for a wide range of parameters values.
An interpretation for this behaviour is that the smallest species progressively adapts to the optimal consumption of resource. Since, in addition, the resource-like species is subject to strong cannibalism, its density and the density of resource become too low for other species to survive.
In Figure~\ref{fig.branstrom.evol_mu.BSC}, a small part of the positive food web remains because the body mass of the smallest species becomes negative before being optimally adapted to the consumption of resources ($z\approx \mu$), hence the remaining resources are sufficient to feed the positive part of the food web.

Replacing the artificial constraints on $z$ and $\mu$ at 0 by a constraint at 1, we obtain Figure~\ref{fig.branstrom.evol_mu.BAC_1}. Contrary to what Figure~\ref{fig.branstrom.evol_mu.BAC} shows, the food web has a non-trivial structure where several species progressively evolve to the constraint: three species have trait $\mu=1$ and one of them has also a log-mass $z=1$.
We emphasise that the artificial constraints that we put in the model in Figures~\ref{fig.branstrom.evol_mu.BAC}-\ref{fig.branstrom.evol_mu.BAC_1} play a key role in the food web evolution since some species reach the boundary of this constraint. Moreover, the sign of the invasion fitness (bottom of Figure~\ref{fig.branstrom.evol_mu}) in the neighbourhood of these species shows that crossing these boundaries should be the natural evolution of the model.

\setlength{\unitlength}{1cm}

\begin{figure}
\captionsetup[subfigure]{justification=centering}
\begin{center}
\begin{subfigure}{0.32\textwidth}
\begin{picture}(0,3.82)
 \put(0,0){\includegraphics[width=5.1cm]{fig3_a1_fw_180705115450_BSC}}
 \put(1.8,0.5){\includegraphics[width=2.5cm]{fig3_a1_fw_tran_180705115450_BSC}}
\end{picture}
\\
\includegraphics[width=5.1cm]{fig3_a2_trait_z_bis_180705115450_BSC}\\
\includegraphics[width=5.1cm]{fig3_a3_trait_mu_bis_180705115450_BSC}\\
 \includegraphics[width=5.1cm]{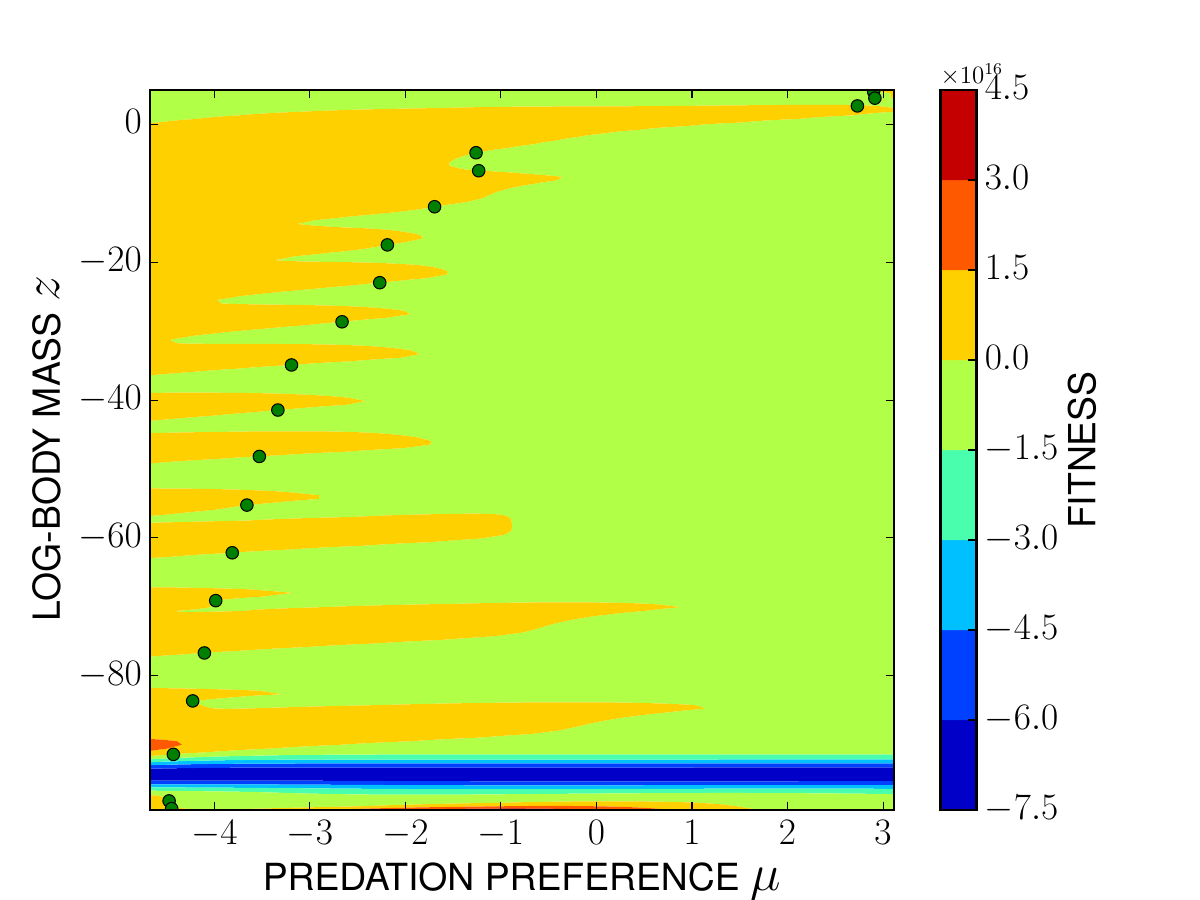}
\caption{\label{fig.branstrom.evol_mu.BSC} without positivity \\ constraints}
\end{subfigure}
\begin{subfigure}{0.32\textwidth}
 \includegraphics[width=5.1cm]{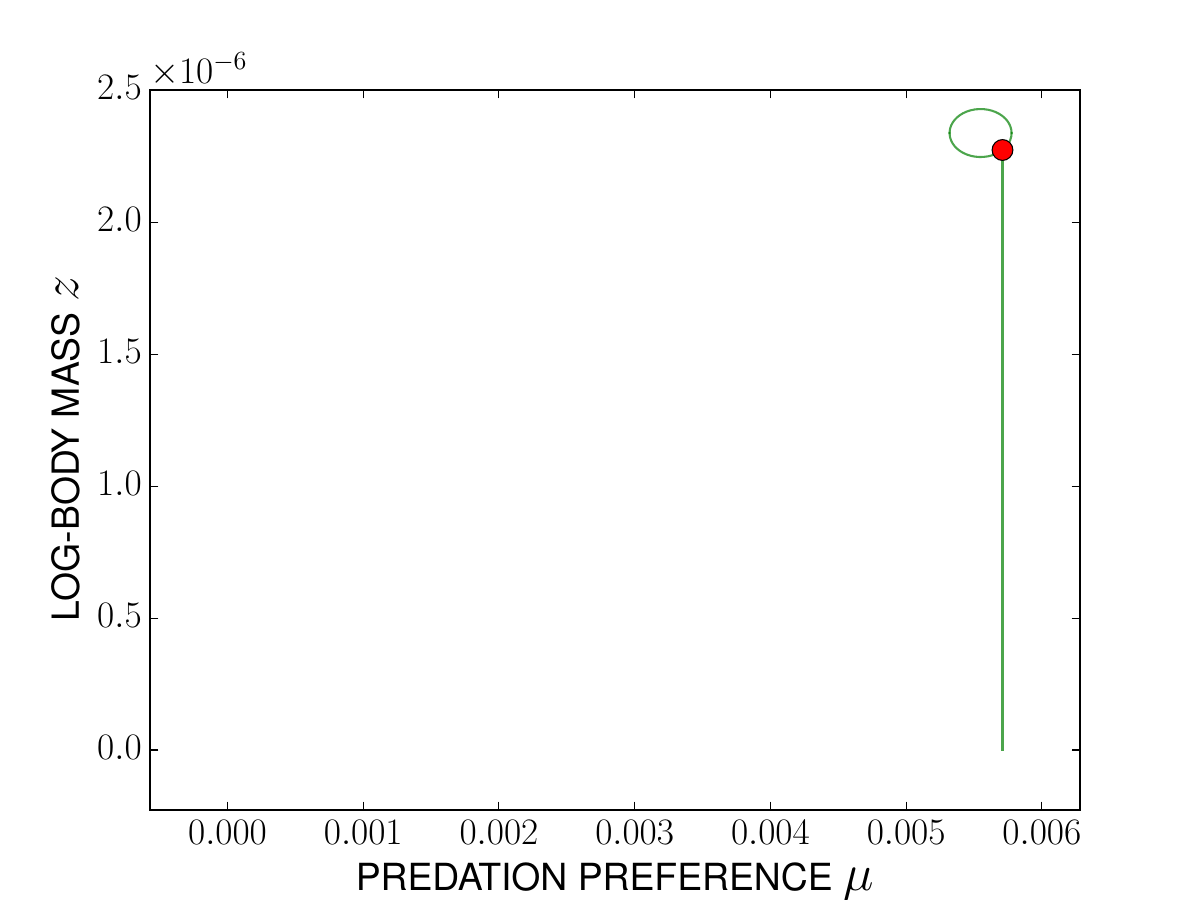}\\
\includegraphics[width=5.1cm]{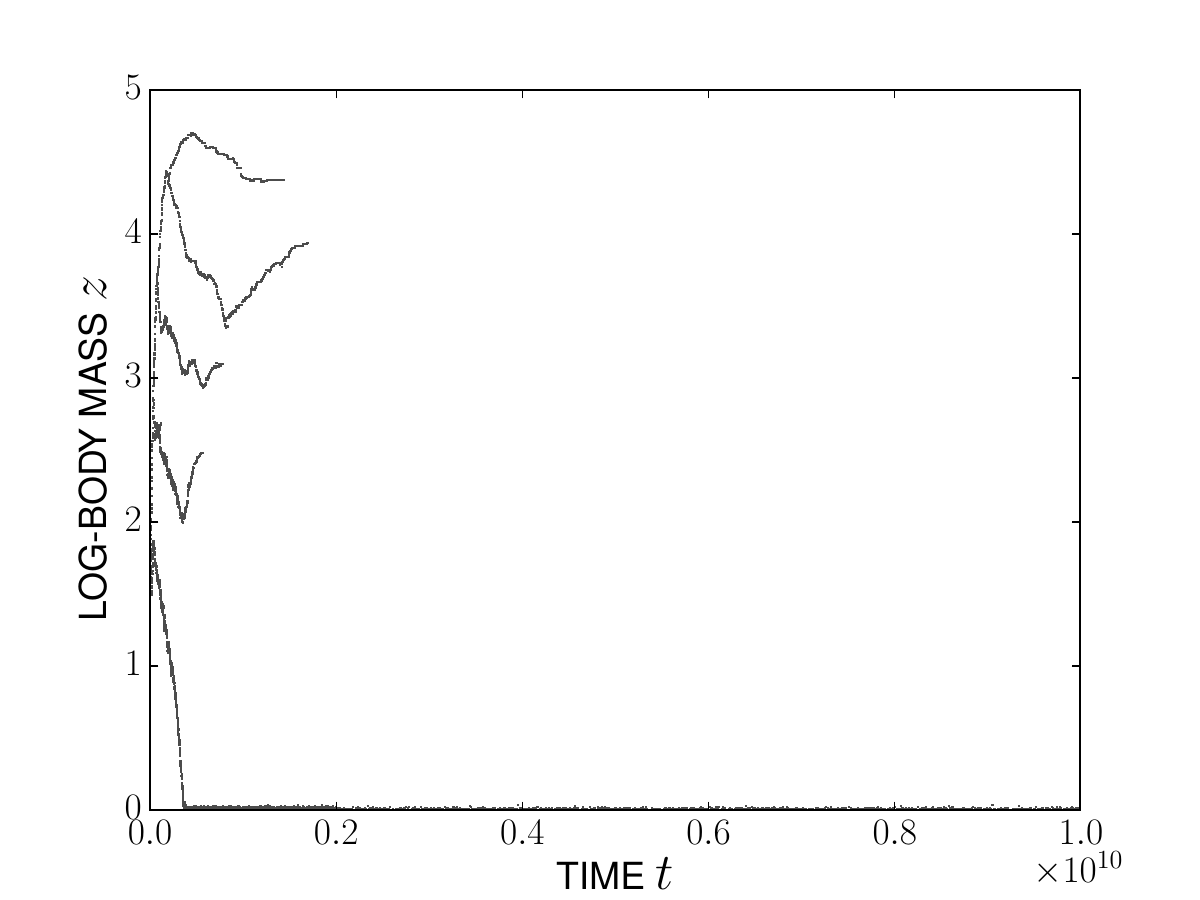}\\
\includegraphics[width=5.1cm]{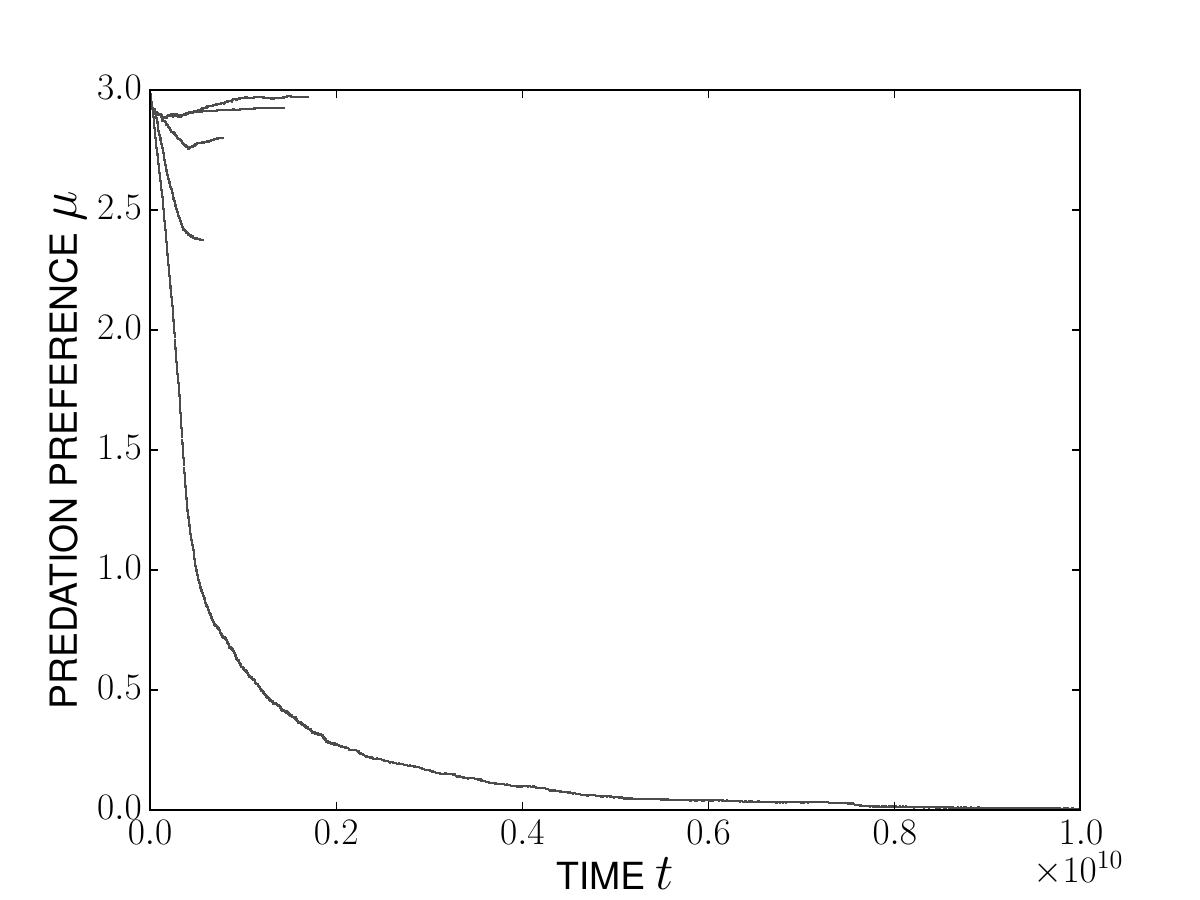}\\
 \includegraphics[width=5.1cm]{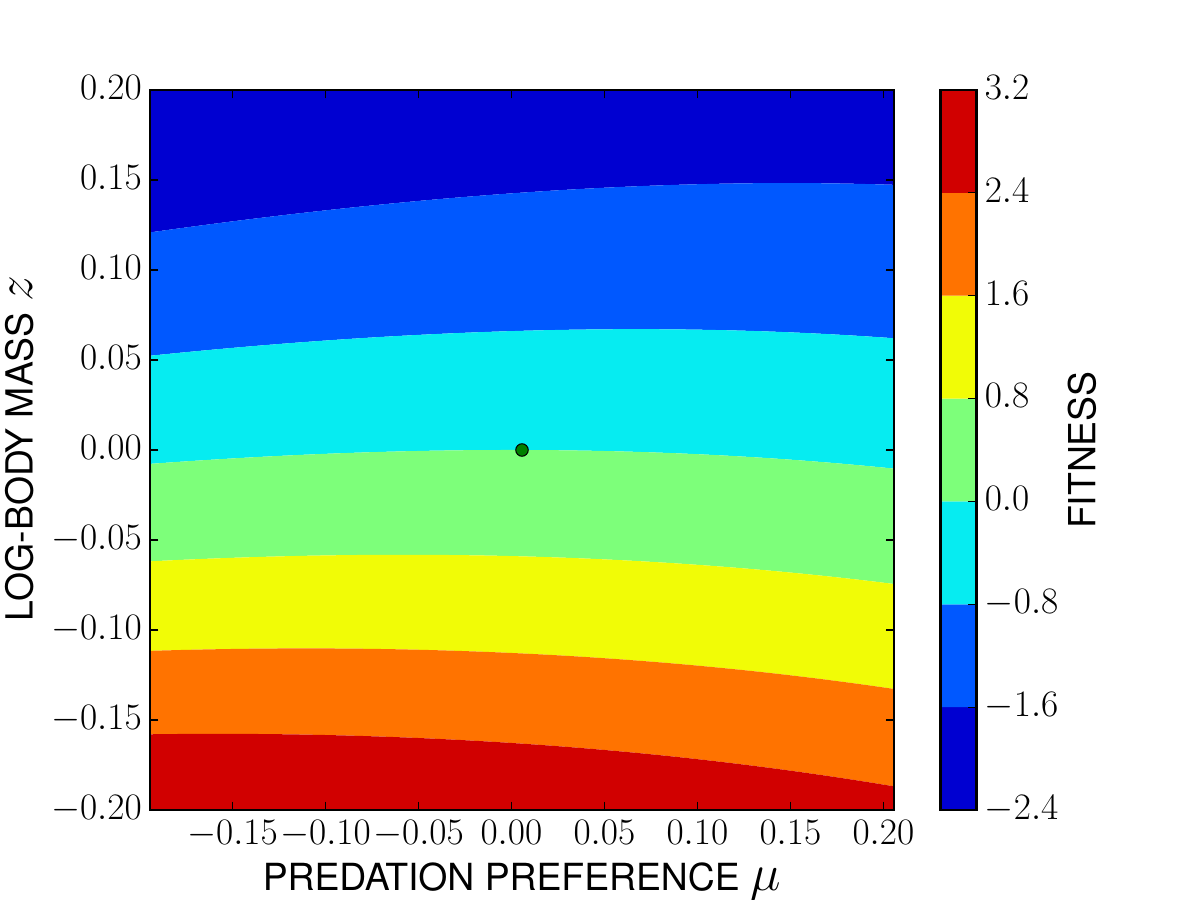}
\caption{\label{fig.branstrom.evol_mu.BAC}  with positivity \\constraints}
\end{subfigure}
\begin{subfigure}{0.32\textwidth}
 \includegraphics[width=5.1cm]{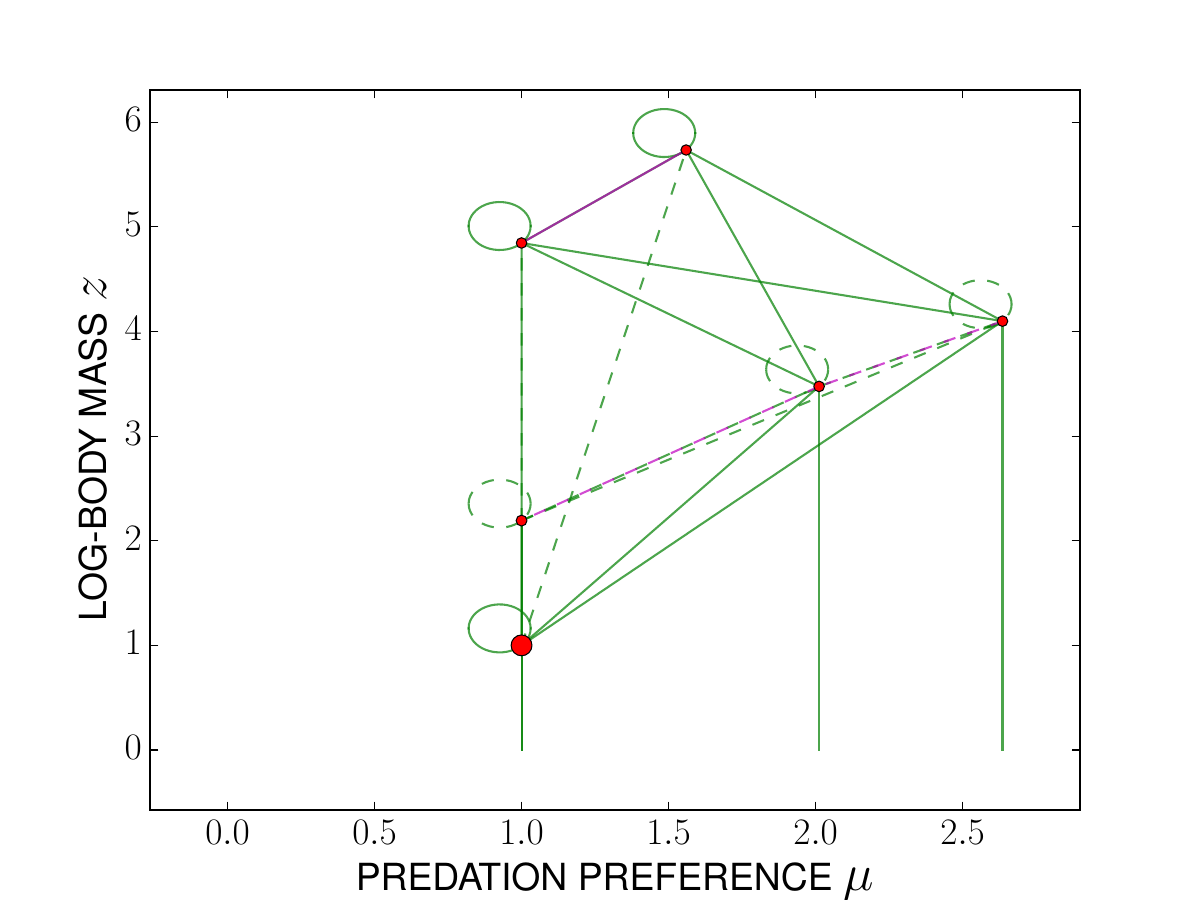}\\
 \includegraphics[width=5.1cm]{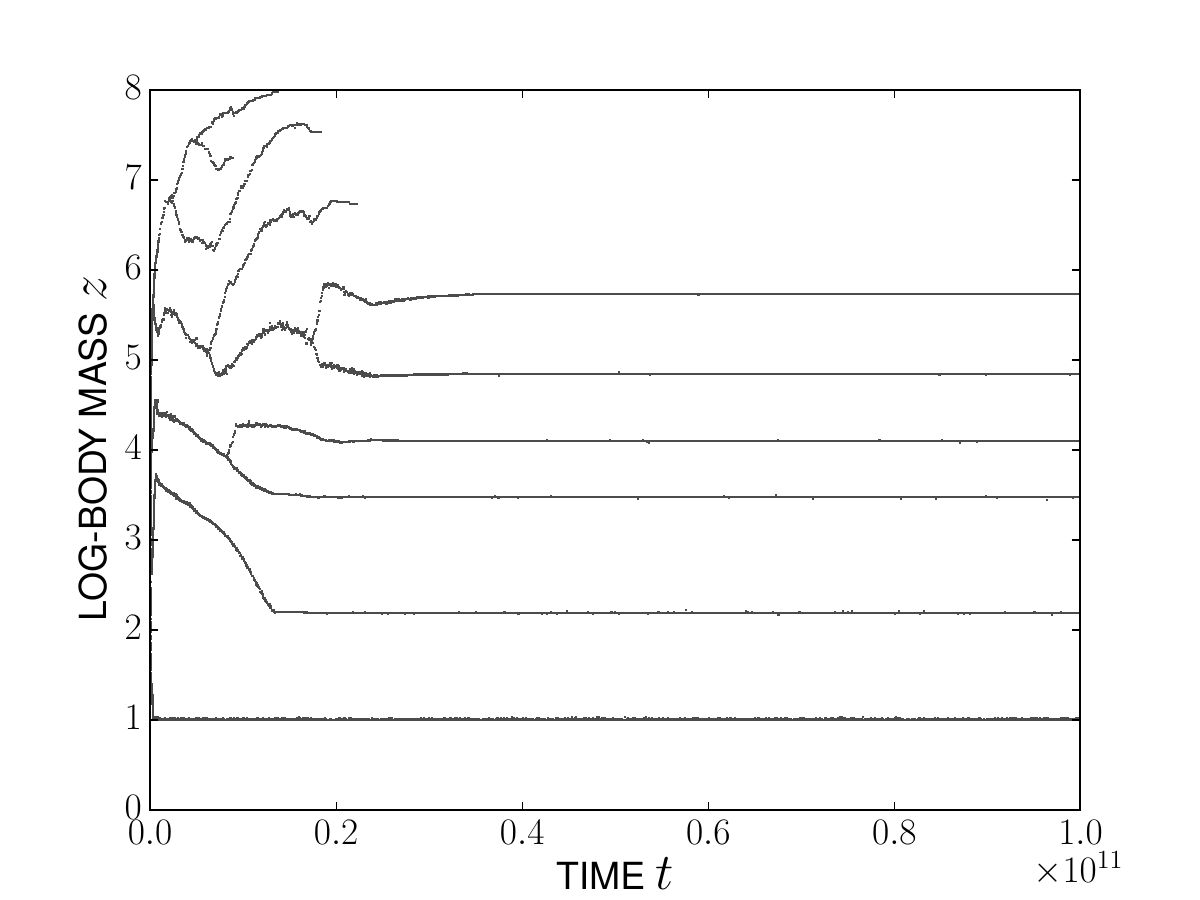}\\
 \includegraphics[width=5.1cm]{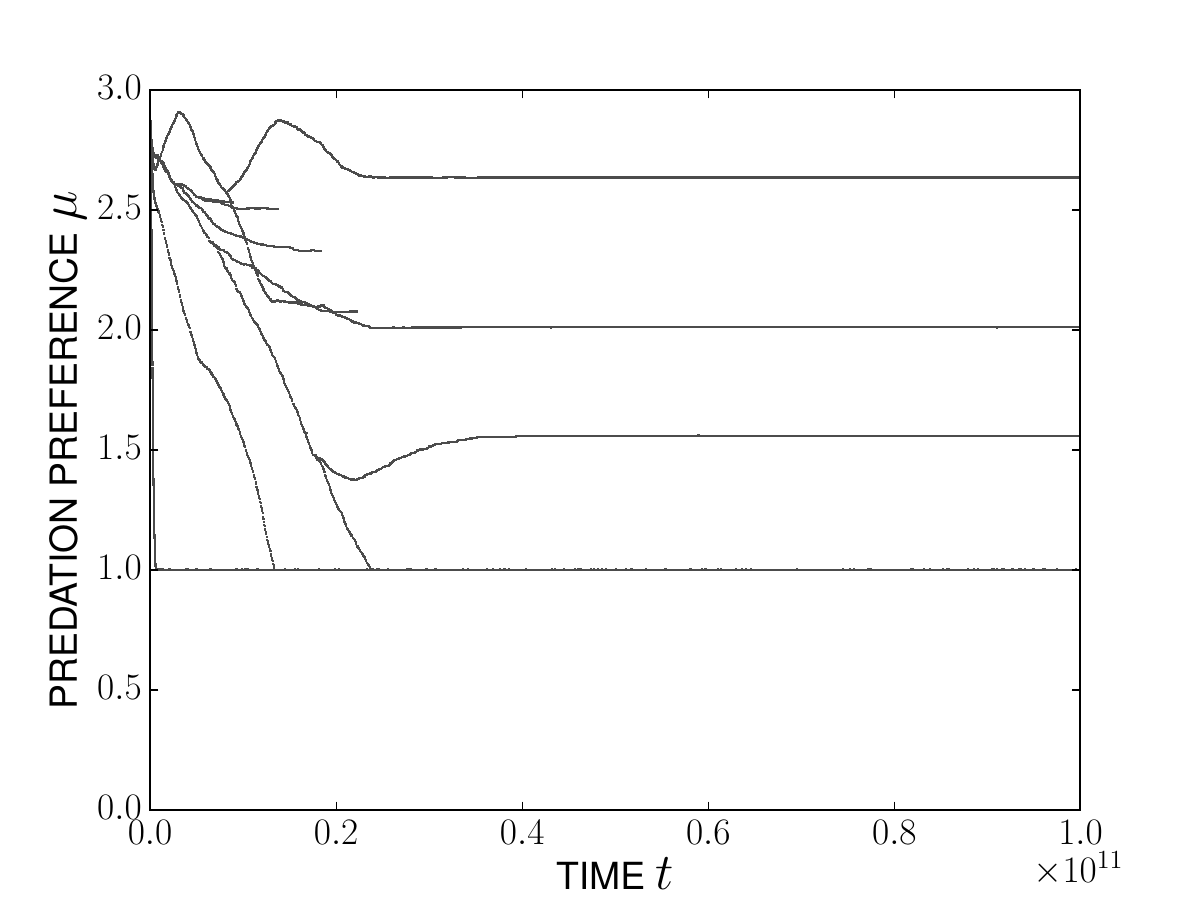}\\
 \includegraphics[width=5.1cm]{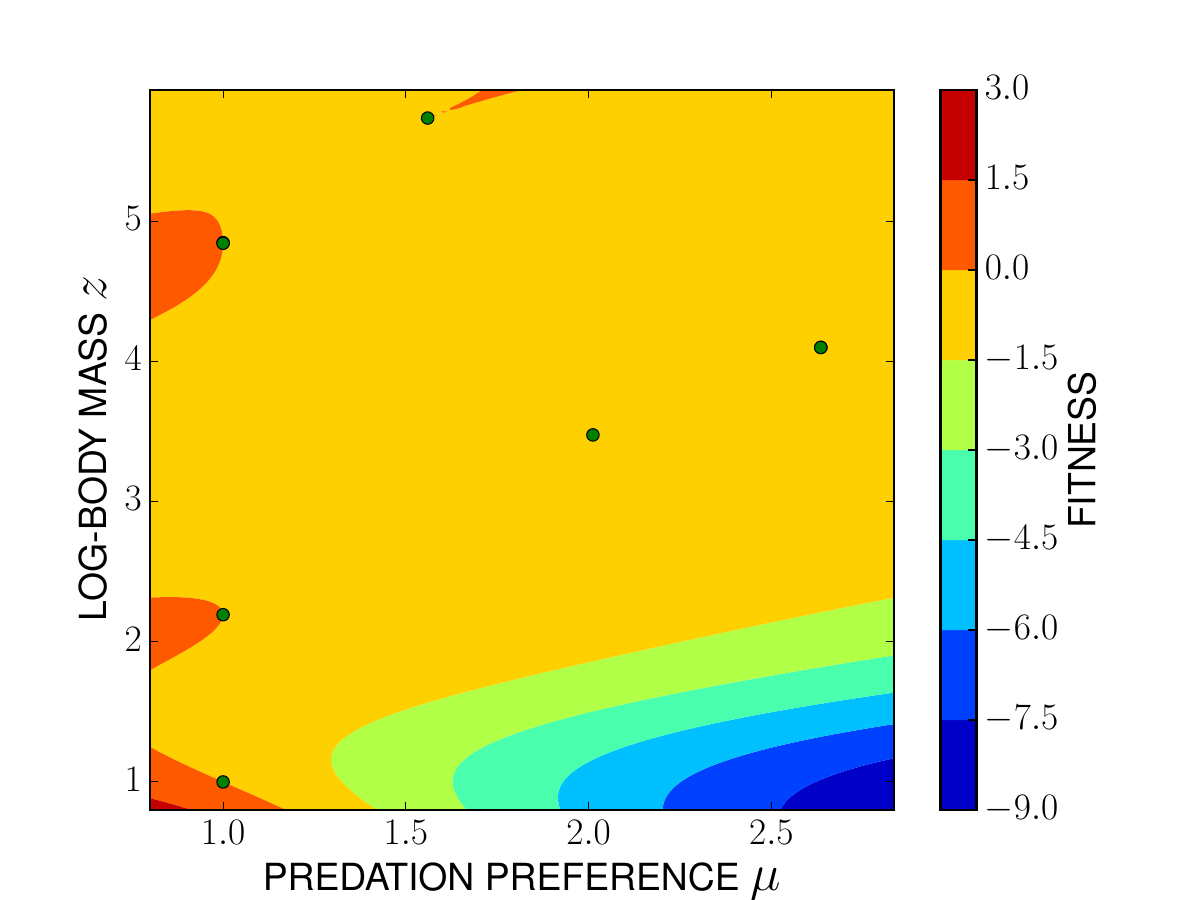}
\caption{\label{fig.branstrom.evol_mu.BAC_1} with constraints \\larger than 1}
\end{subfigure}
\end{center}
\caption{\label{fig.branstrom.evol_mu} Food web evolution relaxing the constraint of fixed predation distance $\mu$. From top to
  bottom: food web at final time of the simulation ($T=2.6\,10^9$, overlay at intermediate time $t=4.\,10^8$ (a); $T=1.\,10^{10}$
  (b); $T=1.\,10^{11}$ (c)); evolution of the log-mass $z$; evolution of the
  predation preference $\mu$; fitness landscape at time $T=2.6\,10^9$ (a);
  $T=1.\,10^{10}$ (b) and $T=1.\,10^{11}$ (c) letting $z$ and $\mu$ evolving in the model of \cite{brannstrom2011} (see
  Table~\ref{table.models} and Section~\ref{sec.model}) without (a), with (b) positivity constraints and with constraints larger than
  1 (c) on evolving traits $z$ and $\mu$ with parameters of Table~\ref{table.parametres}, biomass conversion efficiency $\xi \equiv \lambda_0 = 0.3$, variances of the mutation distributions $\sigma_z=0.01$ and $\sigma_\mu=0.001$.
}
\end{figure}

\medskip

Figure~\ref{fig.branstrom.evol_mu} thus suggests that  \cite{loeuille2005a} and \cite{allhoff2013a} obtained non-trivial food webs only because of artificial and hidden constraints due to the fact that $r_0=0$ and $r_i$ is positive for all $i\geq 1$. In \cite{allhoff2015a} and \cite{allhoff2016a} the constraints on $\mu$ (and $\sigma_\gamma$) are due to the mutation kernels which restrict them to an interval far from zero (see Table~\ref{table.models}). They actually justify this choice because otherwise $\mu$ (and $\sigma_\gamma$) can evolve to arbitrarily small values \citep[see also][]{allhoff2013a}.

Note that other unrealistic patterns were also observed in \cite{allhoff2013a} in the case where only $z$ and $\mu$ evolve: the food web becomes composed mainly of a very large number of species on the first trophic level (with $z\approx \mu$) with a very wide range of body sizes, and a few species on the second trophic level. 
We do not observe this kind of phenomenon, neither in the extension of the model of \cite{brannstrom2011} proposed in this section nor in the one studied in the next section. 
 
Overall, our results suggest that relevant and non-trivial food webs emerge from models derived from \cite{loeuille2005a}  only because of either artificial constraints imposed on the range values of the evolving traits, or because only one trait (size/body mass) evolves (Table \ref{table.models}). In the next section, we propose that relevant food webs could evolve without artificial constraints when considering that biomass conversion is not constant but depends on the relative size of the two interacting species.

\section{First evolutionary branching and the effect of the relative speed of evolution of the two traits}
\label{sec.firstbranching}

We analyse, in this section, the condition under which the first evolutionary branching occurs, {\it i.e.} when a single species coexist with the resource.
We shall use the standard theory of adaptive dynamics to study the first branching event in the food web. We consider a single species $(z,\mu)$ and study the fitness of mutant traits
$(y,\eta)$. The possibility of evolutionary branching is linked to the existence of directions of local convexity of $f(y,\eta)$ in the
  neighbourhood of $(z,\mu)$ \citep{leimar2001a}. In this case, \eqref{eq.f} becomes
\begin{align*}
f(y,\eta) 
	& = 
		\lambda(y,0)\,\gamma(y-\eta)\,N^*_0 + \lambda(y,z)\,\gamma(y-z-\eta)\,N^*_1 
\\
	&\quad - \gamma(z-y-\mu)\,N^*_1- \alpha(y-z)\,N^*_1 - m(y),
\end{align*}
where we deduce from \eqref{eq.equilibre.1pop} that
\begin{equation*}
  N^*_1=\frac{\lambda(z,0)\gamma(z-\mu)\frac{r_g}{k_0}-m(z)}{\frac{\lambda(z,0)\gamma(z-\mu)^2}{k_0}
    +(1-\lambda(z,z))\gamma(-\mu)+\alpha(0)}\quad\text{and}\quad N^*_0=\frac{r_g-\gamma(z-\mu) N^*_1}{k_0}.  
\end{equation*}

\begin{figure}
\begin{center}
 \includegraphics[height=18.7cm, trim = 3.3cm 4.8cm 2.7cm 4.5cm, clip=true]{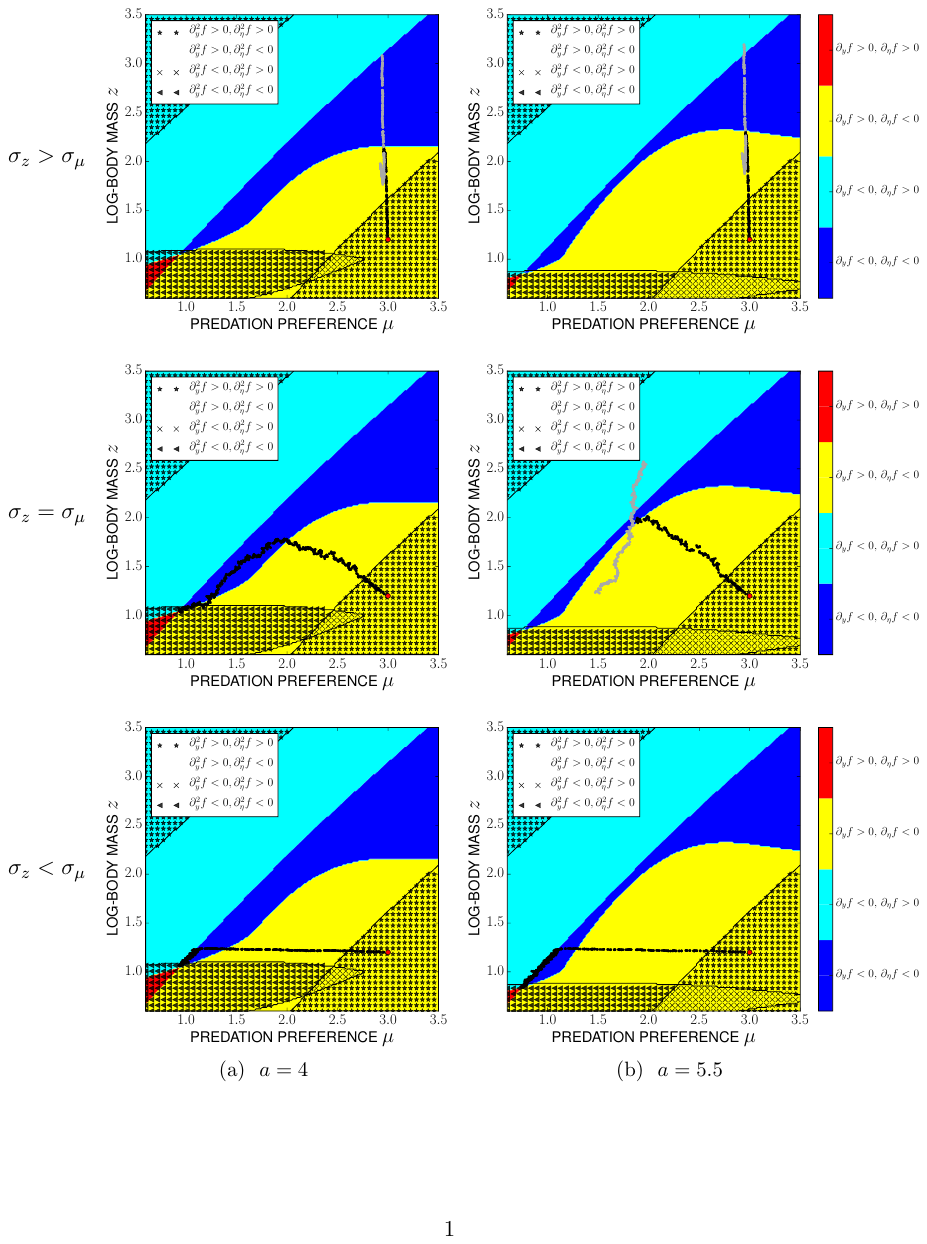}
\end{center}
\vspace{-0.5cm}
\caption{
\label{fig.signs.fitness.derivatives}Sign of the first derivatives \eqref{eq.partial.y.1pop} and \eqref{eq.partial.eta.1pop} and the
second derivatives \eqref{eq.partial2.y.1pop} and \eqref{eq.partial2.eta.1pop} of the invasion fitnesses for $\xi$ defined by \eqref{ex.xi1} with $a=4$ (a) and $a=5.5$
(b) and for mutation variances $0.01=\sigma_z>\sigma_\mu=0.001$, $\sigma_z=\sigma_\mu=0.01$ and $0.001=\sigma_z<\sigma_\mu=0.01$ (from top to bottom).
$b=0.15$ and $\xi_{\max}=0.75$. The red dot corresponds to the starting point of the simulation and the
  black path to the initial evolution of the single species, possibly leading to branching (grey paths correspond to the evolution of both branches).}
\end{figure}

Figure~\ref{fig.signs.fitness.derivatives} show the dynamics of the initial single species for different evolution speeds of both traits (in rows)
and two values of the slope of $\xi$ (in columns) with $\xi$ defined by \eqref{ex.xi1}. In
Figure~\ref{fig.signs.fitness.derivatives} we follow the evolution of both traits of the species. A possible evolutionary pathway is
represented by the successive black dots (and greys dots in case of branching\footnote{Note that after the branching times (for grey dots), the fitness landscapes are different from the ones of Figure~\ref{fig.signs.fitness.derivatives} which are represented for one single species.}) starting from the red dot. The coloured regions give the direction of evolution for both traits (see legend on the right), with respect to the sign of both components of the fitness gradient at the point $(z,\mu)$ given by
\begin{align}
\nonumber
\left. \partial_y f(y,\eta) \right|_{(y,\eta)=(z,\mu)}
	&=
		\left[ \frac{\partial_1 \lambda(z,0)}{\lambda(z,0)}-\frac{z-\mu}{\sigma_\gamma^2}\right]\,
		\lambda(z,0)\,\gamma(z-\mu)\,N_0^*
\\
\nonumber
	&\quad
		+\left[ \frac{\partial_1 \lambda(z,z)}{\lambda(z,z)}+\frac{\mu}{\sigma_\gamma^2}\right]\,
		\lambda(z,z)\,\gamma(-\mu)\,N_1^*
\\
\label{eq.partial.y.1pop}
	&\quad
		+\frac{\mu}{\sigma_\gamma^2}\, \gamma(-\mu)\,N_1^*
		+\frac{m(z)}{4} 
\end{align}
and
\begin{align}
\label{eq.partial.eta.1pop}
\left. \partial_\eta f(y,\eta)\right|_{(y,\eta)=(z,\mu)}
	&=
	 \frac{z-\mu}{\sigma_\gamma^2}\,
		\lambda(z,0)\,\gamma(z-\mu)\,N_0^*
		-\frac{\mu}{\sigma_\gamma^2}\,
		\lambda(z,z)\,\gamma(-\mu)\,N_1^*\,.
\end{align}
For instance, in the yellow region the evolution follows the fitness gradient $\partial_y f>0$ and  $\partial_\eta f<0$ meaning that the log-body size $y$ increases and the preferred predation distance $\eta$ decreases. 
At least in the beginning of the simulations, the initial
species has a tendency to approach the curves $\partial_y f=0$ and $\partial_\eta f=0$ (respectively the frontiers between the yellow and the dark blue region, and between the light blue and dark blue region),
consistently with the classical theory of adaptive dynamics. Note that if $\sigma_z=10\sigma_\mu$ (resp. $\sigma_z=0.1\sigma_\mu$),
evolution is much faster in the $z$ direction (resp. $\mu$ direction) and the population first reaches the line $\partial_y
f=0$ (resp. $\partial_{\eta} f=0$).

The coloured regions are also filled with symbols giving the sign of the second order derivative of fitness, given by
\begin{align}
\nonumber
\left. \partial^2_y f(y,\eta)\right|_{(y,\eta)=(z,\mu)}
	&=
		\left(
		\left[ \frac{\partial_1 \lambda(z,0)}{\lambda(z,0)}-\frac{z-\mu}{\sigma_\gamma^2}\right]^2
		+\frac{\partial^2_1 \lambda(z,0)}{\lambda(z,0)}
		-\left(\frac{\partial_1 \lambda(z,0)}{\lambda(z,0)}\right)^2
		-\frac{1}{\sigma_\gamma^2}\right)\,
		\lambda(z,0)\,\gamma(z-\mu)\,N_0^*
\\
\nonumber
	&\quad
	+\left(
		\left[ \frac{\partial_1 \lambda(z,z)}{\lambda(z,z)}+\frac{\mu}{\sigma_\gamma^2}\right]^2
		+\frac{\partial^2_1 \lambda(z,z)}{\lambda(z,z)}
		-\left(\frac{\partial_1 \lambda(z,z)}{\lambda(z,z)}\right)^2
		-\frac{1}{\sigma_\gamma^2}\right)\,
		\lambda(z,z)\,\gamma(-\mu)\,N_1^*
\\
\label{eq.partial2.y.1pop}
	&\quad
		- \left[\left(\frac{\mu}{\sigma_\gamma^2}\right)^2
		-\frac{1}{\sigma_\gamma^2}\right]\,\gamma(-\mu)\,N_1^*
		+\frac{1}{\sigma_\alpha^2}\alpha(0)\,N_1^* - \frac{m(z)}{16},
\end{align}
\begin{align*}
\nonumber
\left. \partial^2_{y,\eta} f(y,\eta)\right|_{(y,\eta)=(z,\mu)}
	&=
		\left(
		\frac{\partial_1 \lambda(z,0)}{\lambda(z,0)}\,\frac{z-\mu}{\sigma_\gamma^2}-\frac{(z-\mu)^2}{\sigma_\gamma^4}+\frac{1}{\sigma_\gamma^2}\right)\,
		\lambda(z,0)\,\gamma(z-\mu)\,N_0^*
\\
	&\quad
	+		\left(
		-\frac{\partial_1 \lambda(z,z)}{\lambda(z,z)}\,\frac{\mu}{\sigma_\gamma^2}-\frac{\mu^2}{\sigma_\gamma^4}+\frac{1}{\sigma_\gamma^2}\right)\,
		\lambda(z,z)\,\gamma(-\mu)\,N_1^*
\end{align*}
and
\begin{align}
\nonumber
\left. \partial^2_\eta f(y,\eta)\right|_{(y,\eta)=(z,\mu)}
	&=
		\left(
		\left[\frac{z-\mu}{\sigma_\gamma^2}\right]^2
		-\frac{1}{\sigma_\gamma^2}\right)\,
		\lambda(z,0)\,\gamma(z-\mu)\,N_0^*
\\
\label{eq.partial2.eta.1pop}
	&\quad
	+\left(
		\left[\frac{\mu}{\sigma_\gamma^2}\right]^2
		-\frac{1}{\sigma_\gamma^2}\right)\,
		\lambda(z,z)\,\gamma(-\mu)\,N_1^*.
\end{align}
Evolutionary branching is only possible when the second order derivative is positive at the isofitness line. 
As observed in Figure~\ref{fig.signs.fitness.derivatives}, we deduce from these expressions that the curves $\partial_\eta f=0$,
  $\partial^2_{\eta}f=0$ are close to the line $z=\mu$ and the pair of lines $z=\mu\pm\sigma_\gamma$, respectively. This is due to the fact
  that, in the range of parameters we consider, the terms involving $N^*_1$ are negligible with respect to those involving $N^*_0$.
  This implies that $\partial^2_\eta f<0$ whenever $\partial_\eta f=0$, so that evolutionary branching can only occur in the direction $z$ of the trait space.
  
Both for $a=4$ and $a=5.5$, the only evolutionary singularity in the region of traits we consider is located at the bottom left of the pictures (the border point of the red, yellow, light blue and dark blue regions),  where both $\partial^2_y f$ and $\partial^2_\eta f$ are
negative. Hence, this border point is a (local) evolutionary stable strategy where evolutionary branching cannot occur (note that for others values of $a$,  the two regions can overlap such that evolutionary branching can occur in log-body size).
However, for $a=4$ and $\sigma_z>\sigma_\mu$, and $a=5.5$ and $\sigma_z\geq \sigma_\mu$, we
  observe an evolutionary branching in Figure~\ref{fig.signs.fitness.derivatives} which does not take
place at the evolutionary singularity. It takes place along the line
$\partial_y f=0$ at points away from the curve $\partial_\eta f= 0$.
This is explained by evolutionary branching along slow
directional evolution, as described and analysed by \cite{ito2014a}. They claim that such evolutionary branching can occur
along one direction of the trait space when the evolution in the orthogonal directions of the trait space is
slow. The canonical equation of adaptive dynamics \citep{dieckmann1996a, champagnat2001a} predicts that the
speed of evolution in the $\mu$ direction of the trait space is proportional to the fitness gradient $\partial_\eta f$ and the
mutation variance $\sigma^2_\mu$. In cases where $\sigma_\mu=0.1\sigma_z$, evolution in the $\mu$ direction is
  slow. When $\sigma_z=\sigma_\mu$ and $a=5.5$, it is slow enough only  because in this case the branching takes place
at a point close to the line $\partial_\eta f= 0$, hence such that $\partial_\eta f$ is close to zero. 

Overall, the first diversification occurs because of evolutionary branching of the log-body size. Once two morphs with different body size have emerged, both morphs can also evolve diverging preferred predation distance. Moreover the direction of evolution and its diversification strongly depends on the relative speed of the two traits.

\bibliographystyle{apalike}

\end{document}